%% file: paper.tex
\documentclass[11pt]{article}

\newif\iffullver
\fullverfalse
\fullvertrue 
\newcommand{\full}[1]{\iffullver#1\fi}
\newcommand{\short}[1]{\iffullver\else#1\fi}

\usepackage{amsmath, amssymb, amsthm}
\usepackage{algorithm, algorithmic}
\usepackage{verbatim}
\usepackage{xspace}
\usepackage{multirow}
\short{\usepackage{times}}
\short{\usepackage[small,compact]{titlesec}}

\newcommand{\idc}{{iterative database construction}\xspace}
\newcommand{\idcs}{{iterative database constructions}\xspace}
\newcommand{\IDC}{{Iterative Database Construction}\xspace}
\newcommand{\IDCs}{{Iterative Database Constructions}\xspace}
\newcommand{\dus}{database update sequence\xspace}
\newcommand{\DUS}{Database Update Sequence\xspace}
\newcommand{\update}{\mathbf{U}}

\newcommand{\univ}{\mathcal{X}}
\newcommand{\db}{\mathcal{D}}
\newcommand{\dbs}{\mathbb{N}^{|\univ|}}
\newcommand{\dbstruct}{\mathcal{R}_{\update}}

\newcommand{\query}{Q}
\newcommand{\queryset}{\mathcal{Q}}
\newcommand{\queries}{k}

\newcommand{\querystep}[1]{\query^{(#1)}}
\newcommand{\dbstep}[1]{\db^{(#1)}}

\newcommand{\priv}{\eps}
\newcommand{\privd}{\delta}

\newcommand{\threshold}{T}

\newcommand{\noisesize}{\sigma}
\newcommand{\noise}{Z}
\newcommand{\noisestep}[1]{\noise^{(#1)}}
\newcommand{\maxupdates}{B}
\newcommand{\updates}{C}
\newcommand{\Lap}{\mathrm{Lap}}
\newcommand{\acc}{\alpha}
\newcommand{\accf}{\beta}

\newcommand{\real}{A}
\newcommand{\noisyreal}{\widehat{\real}}
\newcommand{\fake}{\Lambda}
\newcommand{\diff}{R}

\newcommand{\realstep}[1]{\real^{(#1)}}
\newcommand{\noisyrealstep}[1]{\noisyreal^{(#1)}}
\newcommand{\fakestep}[1]{\fake^{(#1)}}
\newcommand{\diffstep}[1]{\diff^{(#1)}}

\newcommand{\out}{\mathbf{v}}
\newcommand{\outstep}[1]{\out^{(#1)}}
\newcommand{\outsteps}[1]{\out^{(<#1)}}

\newcommand{\easy}{E_1}
\newcommand{\medium}{E_2}
\newcommand{\hard}{E_3}
\newcommand{\easystep}[1]{\easy^{(#1)}}
\newcommand{\mediumstep}[1]{\medium^{(#1)}}
\newcommand{\hardstep}[1]{\hard^{(#1)}}

\newcommand{\cC}{{\cal C}}

\newcommand{\set}[1]{\left\{#1\right\}}
\newcommand{\card}[1]{\left|\left\{#1\right\}\right|}
\newcommand{\eqdef}{\mathbin{\stackrel{\rm def}{=}}}

\newcommand{\eps}{\epsilon}

\newcommand{\Ex}{\mathbb{E}}

\newcommand{\prob}[1]{\mathrm{Pr}\left[#1\right]}

\newcommand{\R}{\mathbb{R}}
\newcommand{\reals}{\mathbb{R}}

\newtheorem{theorem}{Theorem}[section]
\newtheorem{lemma}[theorem]{Lemma}

\newtheorem{claim}[theorem]{Claim}
\newtheorem{remark}[theorem]{Remark}
\newtheorem{corollary}[theorem]{Corollary}

\theoremstyle{definition}
\newtheorem{definition}[theorem]{Definition}

\newcommand{\initOneLiners}{%
    \setlength{\itemsep}{0pt}
    \setlength{\parsep }{0pt}
    \setlength{\topsep }{0pt}
}
\newenvironment{OneLiners}[1][\ensuremath{\bullet}]
    {\begin{list}
        {#1}
        {\initOneLiners}}
    {\end{list}}

\author{Anupam Gupta\thanks{Department of Computer Science, Carnegie Mellon University, Pittsburgh
    PA 15213. Research was partly supported by
    NSF awards CCF-0964474 and CCF-1016799. Email: {\tt anupamg@cs.cmu.edu}} \and Aaron Roth\thanks{Department of Computer and Information Science, University of Pennsylvania, Philadelphia PA 19104. This research was conducted while at Microsoft Research, New England. Email: {\tt aaroth@cis.upenn.edu}} \and Jonathan Ullman\thanks{School of Engineering and Applied Sciences, Harvard University, Cambridge, MA.  Supported by NSF grant CNS-0831289.  Email: {\tt jullman@seas.harvard.edu}.}}
\title{\bf Iterative Constructions and Private Data Release}

\begin{document}
\maketitle

\input{intro2}

\input{idc}

\input{randomizedresponse}
\input{improvedoffline}

\newpage

\paragraph{Acknowledgements}
This paper benefited from interactions with many people. We particularly
thank Moritz Hardt and Kunal Talwar for extensive, enlightening
discussions. In particular, the observation that randomized response
leads to a data structure for graph cuts with error $O(|V|^{1.5})$ is
due to Kunal Talwar. We thank Salil Vadhan for helpful discussions about
the Frieze/Kannan low-rank matrix decomposition, and Frank McSherry and Adam Smith for helpful discussions about algorithms for computing low-rank matrix approximations.  We thank Cynthia Dwork
for always fruitful conversations.

\bibliographystyle{alpha}
\bibliography{onlinegraphcuts}

\include{appendix}

\end{document}

%% file: intro2.tex
\begin{abstract}

In this paper we study the problem of approximately releasing the \emph{cut function}
of a graph while preserving differential privacy, and give new algorithms
(and new analyses of existing algorithms) in both the interactive
and non-interactive settings.

Our algorithms in the interactive setting are achieved by
revisiting the problem of releasing differentially private, approximate
answers to a large number of queries on a database.  We show
that several algorithms for this problem fall into the same basic framework, and
are based on the existence of objects which we call \emph{iterative database construction}
algorithms.  We give a new generic framework in which new (efficient) IDC algorithms give
rise to new (efficient) interactive private query release mechanisms.  Our modular
analysis simplifies and tightens the analysis of previous algorithms, leading to
improved bounds.  We then give a new IDC algorithm (and therefore a new
private, interactive query release mechanism) based on the Frieze/Kannan low-rank matrix
decomposition.  This new release mechanism gives an improvement on prior work in
a range of parameters where the size of the database is comparable to the
size of the data universe (such as releasing all cut queries on dense graphs).

We also give a non-interactive algorithm for efficiently releasing
private \emph{synthetic data} for graph cuts with error $O(|V|^{1.5})$.
Our algorithm is based on randomized response and a non-private implementation
of the SDP-based, constant-factor approximation algorithm for cut-norm due to
Alon and Naor.  Finally, we give a reduction based on the IDC framework showing
that an efficient, private algorithm for computing sufficiently accurate rank-1 matrix
approximations would lead to an improved efficient algorithm for releasing private
synthetic data for graph cuts.  We leave finding such an algorithm as our main open problem.
\end{abstract}

\short{
\vfill
\thispagestyle{empty}
\setcounter{page}{0}
\pagebreak
}

\section{Introduction}
\label{sec:introduction}

Consider a graph representing the online communications between a set of
individuals: each vertex represents a user, and an edge between two users
indicates that they have corresponded by email.  It might be extremely useful
to allow data analysts access to this graph in order to mine it for statistical information.  However,
the graph is also composed of sensitive information, and we cannot allow
our released information to reveal much about the existence of specific edges.
Thus we would like a way to analyze the structure of this graph while protecting
the privacy of individual edges.  Specifically we would like to be able to provide a promise of \emph{differential privacy} \cite{DMNS06} (defined in Section~\ref{sec:preliminaries}), which, roughly, requires that our algorithms be randomized, and induce nearly the same distribution over outcomes when given two data sets (e.g. graphs) which differ in only a single point (e.g. an edge).

One natural objective is to provide private access to the \emph{cut function}
of this graph. That is, to provide a privacy preserving way for a data
analyst to specify any two (of the exponentially many) subsets of
individuals, and to discover (up to some error) the number of email
correspondences that have passed between these two groups.  There are
two ways we might try to achieve this goal:  We could give an \emph{interactive} solution
where we give the analyst
private oracle access to the cut function.  Here the user can
write down any sequence of cut queries and the oracle will respond with private,
approximate answers.  We may also try for a stronger, \emph{non-interactive} solution, in which we
release a private \emph{synthetic dataset}; a new, private graph that approximately
preserves the cut function of the original graph.

The case of answering cut queries on a graph is just one instance of the more
general problem of query release for exponentially sized families of \emph{linear queries} on
a data set.  Although this problem has been extensively studied in the differential
privacy literature, we observe that no previously known efficient solution is suitable for the case of releasing all
cut queries on graphs.  In this paper we provide solutions to this problem in both
the interactive and non-interactive settings.

We give a generic framework that converts objects
that we call \emph{iterative database construction (IDC)} algorithms into private
query release mechanisms in both the interactive and non-interactive settings.  This
framework generalizes the median mechanism \cite{RR10}, the
online multiplicative weights mechanism \cite{HR10}, and the offline
multiplicative weights mechanism \cite{GHRU11,HLM11}.  Our framework
gives a simple, modular analysis of all of these mechanisms, which lead to
tighter bounds in the interactive setting than those given in \cite{RR10} and \cite{HR10}.
These improved bounds
are crucial to our objective of giving non-trivial approximations to all possible cut queries.
We also instantiate this framework with a new IDC algorithm for arbitrary linear queries that is
based on the Frieze/Kannan
low-rank matrix decomposition~\cite{FK99} and is tailored to releasing cut queries.
This algorithm leads to a new online query release mechanism for linear queries that gives a better
approximation in settings (such as we would encounter trying to answer all cut queries on a dense graph) where the database size is comparable to the size of the
data universe.   We summarize our bounds
in Table~\ref{tab:results}.

We also give a new algorithm (building on techniques for constructing private synthetic data in~\cite{BCDKMT07, DNRRV09}) in the non-interactive setting that efficiently generates private synthetic graphs that approximately preserve the cut function.  Finally, we use our IDC framework to show that an efficient, private
algorithm for the problem of privately computing good rank-1 approximations to symmetric matrices would automatically yield efficient private algorithms for releasing synthetic graphs with improved approximation guarantees.

\begin{table} 
\newcommand\T{\rule{0pt}{3.6ex}}
\newcommand\B{\rule[-2.1ex]{0pt}{0pt}}
\begin{minipage}[center]{\textwidth}
\begin{center}
{\footnotesize
\begin{tabular}{|r|c|c|c|}
\hline
& \multirow{2}{*}{Previous Bounds} & \multicolumn{2}{|c|}{This Paper} \\ \cline{3-4}
&  & General Bounds & Cut Queries \\
\hline
\T \B Median Mechanism\footnote{The bounds listed here are for linear
  queries. The Median Mechanism more generally works for any set of low
  sensitivity queries $\queryset$ that have an $\alpha$-net of size
  $N_\alpha(\queryset)$. We improve the bound from the solution to
  $\alpha = \frac{\log(N_\alpha(\queryset))\log^2(\queryset)}{\epsilon}$
  to the solution to $\alpha = \frac{\sqrt{\log N_\alpha(\queryset)}\log
    \queries}{\epsilon}$.}  \cite{RR10} & $\frac{n^{2/3} (\log
\queries) (\log |\univ|)^{1/3}}{\epsilon^{1/3}}$ \cite{RR10} &
$\frac{n^{1/2} (\log \queries)^{3/4}(\log |\univ|)^{1/4}}{\epsilon^{1/2}}$ & $\frac{|E|^{1/2}|V|^{3/4}(\log|V|)^{1/4}}{\epsilon^{1/2}}$
  \\
\hline
\T \B Online MW \cite{HR10} & $\frac{ n^{1/2} (\log \queries ) (\log|\univ|)^{1/4}}{\epsilon}$ \cite{HR10}& $\frac{n^{1/2}(\log \queries)^{1/2}(\log |\univ|)^{1/4}}{\epsilon^{1/2}}$ & $\frac{|E|^{1/2}|V|^{1/2}(\log|V|)^{1/4}}{\epsilon^{1/2}}$ \\ \hline
\T \B Frieze/Kannan IDC & New in this paper & $\frac{n_2^{1/4}(\log
\queries)^{1/2}|\univ|^{1/4}}{\epsilon^{1/2}}~~~\footnote{Here we use $n_2 =
  \|\db\|_2^2$, in contrast to other known IDCs, whose error is in terms
  of $n = \|\db\|_1$.  Note that $n \leq n_2 \leq n^2$.}$ & $\frac{|E|^{1/4}|V|}{\epsilon^{1/2}}$ \\ \hline
\T \B K-Norm Mechanism \cite{HT10} &
$\frac{\sqrt{\queries}}{\epsilon}\left(\log\left(\frac{|\univ|}{\queries}\right)\right)^{1/2}$\cite{HT10}\footnote{For
  $\queries \leq |\univ|/2$. This is an approximate bound on
  \emph{average} per-query error. All other algorithms listed bound
  worst-case per-query error.} &  Not in the IDC Framework &
Not Applicable \\ \hline
\end{tabular}
}
\caption{\label{tab:results} Comparison of accuracy bounds for linear queries. The bounds in the first column are prior to this work, the second column are what we achieve in this work, and the last column are the new bounds instantiated for releasing all cut queries. The bounds listed here are approximate and hide the dependence on certain parameters, such as $\delta$ and $\beta$. $n$ denotes database size, $k$ denotes the total number of queries answered, and $\univ$ represents the data universe. For a graph $G = (V,E)$, $n = n_2 = |E|$, $|\univ| = {|V| \choose 2}$, and for all cut queries, $k = 2^{2|V|}$. Previous efficient results do not achieve non-trivial ($\leq |E|$) error, while all of the new bounds do for sufficiently dense graphs.}
\end{center}
\end{minipage}
\end{table}

\subsection{Our Results and Techniques}
\label{sec:techniques}

Our main conceptual contribution is to define the abstraction
of \emph{iterative database construction} algorithms (Section~\ref{sec:idc}) and
to show that an efficient IDC for any class of queries $\queryset$ automatically
yields an efficient private data release mechanism for $\queryset$ in both the
interactive and non-interactive settings.  Informally, IDCs construct a data structure
that can be used to answer all the queries in $\queryset$ by iteratively improving
a hypothesis data structure.  Moreover, they update the hypothesis when given a query
witnessing a significant difference between the hypothesis data structure and the
underlying database.

In hindsight, this framework generalizes the median mechanism \cite{RR10} and the
subsequent refinement for linear queries, the online multiplicative weights mechanism \cite{HR10}.
It also generalizes the offline multiplicative weights mechanism \cite{GHRU11, HLM11}.
All of these mechanisms can be seen to use IDCs of the sort we define in
this work.  (In Appendix \ref{app:other-IDCs} we show how these
algorithms fall into the IDC framework.)

Our generalization and abstraction also allows for a simple, modular analysis of
mechanisms based on IDCs.  Using this analysis, we are able to show improved bounds
on the accuracy of both the median mechanism and multiplicative weights mechanism.
These improved bounds are critical to our application to releasing all cut queries.  For these
parameters, the previous bounds would not guarantee error that is $\leq |E|$, meaning
that the error may be larger than the largest cut in the graph.  Of course, we can privately
guarantee error $\leq |E|$ simply by releasing the answer $0$ for every cut query.  Our
new analysis shows that these mechanism are capable of answering all cut queries with
error $o(|E|)$ for sufficiently dense graphs.

We also define a new IDC based on the Frieze/Kannan low-rank matrix decomposition~\cite{FK99},
which yields a private interactive mechanism for releasing linear queries.  Our new mechanism
outperforms previously known techniques when the size of the database is comparable
to the size of the data universe, as is the case on a dense graph.

We then consider the problem of efficiently releasing private synthetic data
for the class of cut queries.  We show that a technique based on randomized
response efficiently yields a private data structure (but not a synthetic databse)
capable of answering any cut query on a graph with $|V|$ vertices up to
maximum error $O(|V|^{1.5})$.  We then show how to use this data structure
to efficiently construct a synthetic database with only a constant factor blowup in our
error.  Our algorithm is based on a technique for constructing synthetic data in \cite{BCDKMT07, DNRRV09}.
Their observation is that, for linear queries, the set of accurate synthetic databases is described
by a (large) set of linear constraints.  In the case of cut queries, we are able to use
a constant-factor approximation to the cut-norm due to
Alon and Naor \cite{AN03} as the
separation oracle to find a feasible solution (and thus a synthetic database)
efficiently.  Finally, we show how the
existence of an efficient private algorithm for finding good low-rank
approximations to matrices would imply the existence of an improved
algorithm for privately releasing synthetic data for cut queries, using our IDC framework.

\subsection{Related Work}
\label{sec:related-work}

Differential privacy was introduced in a series of papers
\cite{BDMN05,CDMSW05, DMNS06} in the last decade, and has become a
standard solution concept for statistical database privacy.  The first mechanism
for simultaneously releasing the answers to exponentially large classes
of statistical queries was given in \cite{BLR08}.  They
showed that the existence of small nets for a class of queries $\queryset$
automatically yields a (computationally inefficient) non-interactive, private algorithm
for releasing answers to all the queries in $\queryset$ with low error.
Subsequent improvements were given by Dwork et al. \cite{DNRRV09,DRV10}.

Roth and Roughgarden \cite{RR10} showed that large classes of
queries could also be released with low error in the
\emph{interactive} setting, in which queries may arrive online, and the
mechanism must provide answers before knowing which queries will arrive
in the future. Subsequently, Hardt and Rothblum \cite{HR10} gave
improved bounds for the online query release problem based on the
multiplicative weights algorithm. In hindsight, both of these algorithms follow the
same basic framework, which is to use an IDC.

Gupta et al.~\cite{GHRU11} gave a \emph{non-interactive} data release
mechanism based on the multiplicative weights algorithm and an arbitrary
agnostic learner for a class of queries. An instantiation of this
algorithm (the \emph{offline} multiplicative weights algorithm) using
the generic agnostic learner of Kasiviswanathan et al. \cite{KLNRS08}
(who use the exponential mechanism of \cite{MT07}) was implemented and
experimentally evaluated on the task of releasing small conjunctions to
low error on real data by Hardt, Ligett, and McSherry \cite{HLM11}. This
algorithm gives bounds comparable to those given in this paper, but it
does not work in the interactive setting, and is not computationally
efficient for settings in which the number of queries is exponentially larger than the
database size (as is the case with graph cuts).  We note in Section
\ref{sec:improvedoffline} that this generic algorithm can also be
instantiated with any iterative database construction algorithm.

Hardt and Talwar \cite{HT10} consider the setting where the number of queries is smaller
than the universe size. When the number of queries
is comparable to the universe size (i.e. $|\queryset| = \Omega(|\univ|)$),
their K-Norm mechanism gives \emph{average error} that is smaller
than the worst-case error promised by the online multiplicative weights
mechanism when the database size is $n \geq \tilde{O}\left(\frac{|\univ|}{|\log|\univ|}\right)$.
This is the same
range of parameters for which the Frieze/Kannan IDC algorithm improves
on the online-multiplicative weights, and in this range of parameters, it
achieves roughly the same error as the K-norm mechanism.  In general,
the bounds for the two mechanisms are incomparable: e.g., \cite{HT10}
have a better, logarithmic dependence on $|\univ|$, compared to the
polynomial dependence for the Frieze/Kannan IDC. On the other hand,
the Frieze/Kannan IDC (and all algorithms in the IDC framework)
have some advantages.  Specifically, the bounds are for worst-case error, rather
than average-case error; hold unconditionally, while
the accuracy of the K-norm mechanism relies on the truth of the hyperplane conjecture;
apply even when the number of queries is larger than the universe size; and
typically have running time linear in $|\univ|$, rather than $\mathrm{poly}(|\univ|)$.

The Frieze-Kannan low-rank approximation (or the weak regularity lemma)
shows that every matrix can be approximated by a sum of few cut
matrices~\cite{FK99,FK-svd}: this fact has many important algorithmic
applications. We also use the fact that the proof extends to more general
settings, as was noted by~\cite{TTV09}.

\section{Preliminaries}
\label{sec:preliminaries}

In this paper, we study datasets $\db$ that consist of collections of
$n$ elements from some universe $\univ$. We can also write $\db \in
\dbs$ when it is convenient to represent $\db$ as a histogram over
$\univ$. We say that two databases $\db$, $\db'$ are adjacent if they
differ in only a single element. As histograms, they are adjacent if
$\|\db-\db'\|_1 \leq 1$. We will require that our algorithms satisfy
\emph{differential privacy}:
\begin{definition}[Differential Privacy]
  A randomized algorithm $M:\dbs\rightarrow R$ (for any abstract range
  $R$) satisfies $(\epsilon,\delta)$-differential privacy if for all
  adjacent databases $\db$ and $\db'$, and for all events $S \subseteq
  R$: \short{$\Pr[M(\db) \in S] \leq \exp(\epsilon)\Pr[M(\db') \in S] +\delta$}
  \full{$$\Pr[M(\db) \in S] \leq \exp(\epsilon)\Pr[M(\db') \in S] +\delta$$}
\end{definition}
We will generally think of $\epsilon$ as being a small constant, and
$\delta$ as being negligibly small -- i.e. smaller than any inverse
polynomial function of $n$.

We note that when we will discuss interactive mechanisms, we must view the output of a mechanism as a \emph{transcript} of an interaction between an adaptive adversary who supplies questions about the database based on previous outcomes of the mechanism, and the mechanism itself. For clarity, in this paper we will elide specifics about the model of adaptive private composition. For a detailed treatment of this issue, see \cite{DRV10}.

A useful distribution is the \emph{Laplace} distribution.
\begin{definition}[The Laplace Distribution]
The Laplace Distribution (centered at 0) with scale $b$ is the
distribution with probability density function: $\textstyle \Lap(x | b)
= \frac{1}{2b}\exp( -\frac{|x|}{b})$.
We will sometimes write $\textrm{Lap}(b)$ to denote the Laplace distribution with scale $b$, and will sometimes abuse notation and write $\Lap(b)$ simply to denote a random variable $X \sim \Lap(b)$.
\end{definition}
A fundamental result in data privacy is that perturbing low sensitivity queries with Laplace noise preserves $(\epsilon,0)$-differential privacy.
\begin{theorem}[\cite{DMNS06}]
\label{thm:laplace-privacy}
Suppose $\query:\dbs\rightarrow \R^k$ is a function such that for all
adjacent databases $\db$ and $\db'$, $\|\query(\db) - \query(\db')\|_1
\leq 1$. Then the procedure which on input $\db$ releases $\query(\db) +
(X_1,\ldots,X_k)$, where each $X_i$ is an independent draw from a
$\textrm{Lap}(1/\epsilon)$ distribution, preserves
$(\epsilon,0)$-differential privacy.
\end{theorem}
It will be useful to understand how privacy
parameters for individual steps of an algorithm compose into privacy
guarantees for the entire algorithm.  The following useful theorem is due to
Dwork, Rothblum, and Vadhan:
\begin{theorem}[\cite{DRV10}]\label{thm:expectedprivacyloss}
Let $0 \leq \priv \leq 1$ be a parameter.  Let $P, Q$ be probability
measures supported on a set $\mathcal{S}$ such that
$\max_{s \in S} \left| \log\left(P(s)/Q(s)\right) \right| \leq \priv.$  Then
$\Ex_{P} \left[ \log\left(P(s)/Q(s)\right) \right] \leq 2\priv^2.$
\end{theorem}


We are interested in privately releasing accurate answers to large
collections of queries. Queries are functions $\query:\dbs\rightarrow
\R$, and we denote collections of queries by $\queryset$. We write $k =
|\queryset|$ to denote the cardinality of the set of queries.

A common type of queries are \emph{linear queries}. A linear query $\query$ has a representation as a vector $[0,1]^{|\univ|}$, and can be evaluated on a database by taking the dot product between the query and the histogram representation of the database: $\query(\db) = \query \cdot \db$.

\begin{definition}[Accuracy]\label{def:accuracy}
  Let $\queryset$ be a set of queries.
  A mechanism $M: \dbs \to \mathcal{R}$ is \emph{$(\alpha,\beta)$-accurate for $\queryset$}
  if there exists a function $\mathrm{Eval}: \queryset \times \mathcal{R} \to \mathbb{R}$ s.t.
  for every database $\db \in \dbs$, with
  probability at least $1-\beta$ over the coins of $M$, $M(\db)$ outputs $r \in \mathcal{R}$
  such that $\max_{\query \in \queryset}|\query(\db)- \mathrm{Eval}(\query, r)| \leq \alpha$.
  We will abuse notation and write $\query(r) = \mathrm{Eval}(\query, r)$.
\end{definition}

We say that an algorithm $M$ \emph{releases synthetic data} (as is the
case for our new IDC, as well as the multiplicative weights
IDC~\cite{HR10}) if $\mathcal{R} = \dbs$ In this case, $M(\db) = \db'
\in \dbs$ and $\mathrm{Eval}(\db', \query) = \query(\db')$.  We say that
a synthetic data release algorithm is \emph{efficient} if it runs in
time polynomial in $n = \|\db\|_1$, the size of the data set. Note that
if $n \ll |\univ|$, efficient algorithms will have to input and output
concise representations of the dataset (i.e., as collections of items
from the universe) instead of using the histogram
representation. Nevertheless, it will be convenient to think of datasets
as histograms.

We say an algorithm efficiently releases $k$ queries from a class
$\queryset$ in the interactive setting if on an arbitrary, adaptively
chosen stream of queries $\query_1,\ldots,\query_k$, it outputs answers
$a_1,\ldots,a_k$. The algorithm must output each $a_i$ after receiving
query $\query_i$ but before receiving $\query_{i+1}$, and is only
allowed poly$(n)$ run time per query. We are typically interested in the
case when $k$ can be exponentially large in $n$. Note that as far as
computational efficiency is concerned, releasing synthetic data for a
class of queries $\queries$ is at least as difficult as releasing
queries from $\queries$ in the interactive setting, since we can use the
synthetic data to answer queries interactively.


\medskip
\noindent\textbf{Graphs and Cuts.}
When we consider datasets that represent graphs $G = (V, E)$, we think
of the database as being the edge set $\db_G = E$, and the data-universe
being the collection of all possible edges in the complete graph:
$|\univ| = {|V| \choose 2}$. That is, we consider the vertex set to be
common among all graphs, which differ only in their edge sets. One
example we care about is approximating the cut function of a
private graph~$G$.

For any real-valued matrix $A \in \R^{m \times m'}$, for $S \subseteq
[m]$ and $T \subseteq [m']$, we define $A(S,T) := \sum_{s \in S, t \in
  T} A_{st}$.  The \emph{cut norm of the matrix} $A$ is now defined as
$\|A\|_C := \max_{S \subseteq [m], T \subseteq [m']} |A(S,T)|$. A graph
$G$ can be represented as its adjacency matrix $A_G \in
\{0,1\}^{|V|\times|V|}$. In this paper, a \emph{cut} in a graph $G$ is
defined by any two subsets of vertices $S,T \subseteq V$. We write the
value of an $S,T$ cut in $G$ as $G(S,T) := A_G(S,T)$, where $A_G$ is the
adjacency matrix of $G$. Similarly, we extend the definition of
\emph{cut norm} to $n$ vertex graphs naturally by defining
$\|G\|_C := \|A_G\|_C = \max_{S, T \subseteq V} \left|G(S,T)\right|$ and
$\|G - H\|_C := \|A_G - A_H\|_C$.
The class of cut queries $\queryset_{\textrm{Cut}} = \{Q_{S,T} : S, T \subseteq V\}$, where $Q_{S,T}(G) = A_G(S,T)$. Note that cut queries are an example of a class of linear queries, because we can represent them as a vector in which $Q_{S,T}[i,j] = 1$ if $i \in S, j \in T$ and $0$ otherwise, and evaluate $Q_{S,T}(G) = \sum_{i,j \in V}Q_{S,T}[i,j]\cdot A_G[i,j]$.

Note that as linear queries, we can write cut queries as the outer product of two vectors: $Q_{S,T} = \chi_S\cdot\chi_T^T$, where $\chi_S,\chi_T \in \{0,1\}^{|V|}$ are the characteristic vectors of the sets $S$ and $T$ respectively. Let us define a more general class of \emph{rank-1} queries on graphs to be a subset of all linear queries: \short{$\queryset_{r1} = \{\query \in [0,1]^{|V|\times|V|}\ \textrm{such that }Q = u\cdot v^T\ \textrm{for some vectors } u,v \in [0,1]^{|V|}\}$}
\full{$$\queryset_{r1} = \{\query \in [0,1]^{|V|\times|V|}\ \textrm{such that }Q = u\cdot v^T\ \textrm{for some vectors } u,v \in [0,1]^{|V|}\}$$}
\full{ A rank-1 query is a linear query and can be evaluated
$$\query_{u,v}(G) = \sum_{i,j \in V}\query[i,j]A_G[i,j] = \sum_{i,j \in V}u[i]v[j]A_G[i,j]$$}
Of course the set of rank-1 queries includes the set of cut queries, and any mechanism that is accurate with respect to rank-1 queries is also accurate with respect to cut queries.



%% file: idc.tex
\section{\IDCs}
\label{sec:idc}

In this section we define the abstraction of \emph{\idcs}that includes our new Frieze/Kannan construction and several existing algorithm \cite{RR10, HR10} as a special case.  Roughly, each of these mechanisms works by maintaining a sequence of data structures $\dbstep{1}, \dbstep{2}, \dots$ that give increasingly good approximations to the input database $\db$ (in a sense that depends on the IDC).  Moreover, these mechanisms produce the next data structure in the sequence by considering only one query $\query$ that \emph{distinguishes} the real database in the sense that $\query(\dbstep{t})$ differs significantly from $\query(\db)$.

Syntactically, we will consider functions of the form $\update: \dbstruct \times \queryset \times \R \to \dbstruct$.  The inputs to $\update$ are a data structure in $\dbstruct$, which represents the current data structure $\dbstep{t}$; a query $\query$, which represents the distinguishing query, and may be restricted to a certain set $\queryset$; and also a real number. which estimates $\query(\db)$.  Formally, we define a \emph{\dus}, to capture the sequence of inputs to $\update$ used to generate the database sequence $\dbstep{1}, \dbstep{2}, \dots$.

\begin{definition}[\DUS]~\label{def:dus}
Let $\db \in \dbs$ be any database and let \\ $\set{(\dbstep{t},
  \querystep{t}, \noisyrealstep{t})}_{t=1,\dots, \updates} \in
(\dbstruct \times \queryset \times \R)^{\updates}$ be a sequence of
tuples.   We say the sequence is an \emph{$(\update, \db, \queryset,
  \acc, \updates)$-\dus} if it satisfies the following properties:
\medskip
\begin{OneLiners}
\item[1.] $\dbstep{1} = \db(\emptyset, \cdot, \cdot)$,
\item[2.] for every $t = 1,2,\dots,\updates$, $\left| \querystep{t}(\db) - \querystep{t}(\dbstep{t}) \right| \geq \acc$,
\item[3.] for every $t = 1,2,\dots,\updates$, $\left| \querystep{t}(\db) - \noisyrealstep{t} \right| < \acc$,
\item[4.] and for every $t = 1,2, \dots, \updates-1$, $\dbstep{t+1} = \update(\dbstep{t}, \querystep{t}, \noisyrealstep{t})$.
\end{OneLiners}
\end{definition}
We note that for all of the \idcs we consider, the approximate answer $\noisyrealstep{t}$ is used only to determine the \emph{sign} of $ \querystep{t}(\db) - \querystep{t}(\dbstep{t})$, which is the motivation for requiring that $\noisyrealstep{t}$ have error smaller than $\acc$.  The main measure of efficiency we're interested in from an \idc is the maximum number of updates we need to perform before the database $\dbstep{t}$ approximates $\db$ well with respect to the queries in $\queryset$.  To this end we define an \idc as follows:

\begin{definition}[\IDC]~\label{def:idc}
Let $\update: \dbstruct \times \queryset \times \R \to \dbstruct$ be an update rule and let $\maxupdates: \R \to \R$ be a function.  We say $\update$ is a \emph{$\maxupdates(\acc)$-\idc for query class $\queryset$} if for every database $\db \in \dbs$, every $(\update, \db, \queryset, \acc, \updates)$-\dus satisfies $\updates \leq \maxupdates(\acc)$.
\end{definition}

Note that the definition of an $\maxupdates(\acc)$-\idc implies that if $\update$ is a $\maxupdates(\acc)$-\idc, then given any  maximal $(\update, \db, \queryset, \acc, \updates)$-\dus, the final database $\dbstep{\updates}$ must satisfy
$\max_{\query \in \queryset} \left| \query(\db) - \query(\dbstep{\updates}) \right| \leq \acc$
or else there would exist another query satisfying property 2 of Definition~\ref{def:dus}, and thus there would exist a $(\update, \db, \queryset, \acc, \updates+1)$-\dus, contradicting maximality.

\section{Query Release from Iterative Database Construction}

In this section we describe an interactive algorithm for releasing linear queries using an arbitrary \idc.

\begin{algorithm}
\noindent\textbf{$\mathcal{M}^{\update}(\db, \priv, \privd, \acc, \accf, \queries)$:}
\begin{algorithmic}
\STATE{\textbf{Input:} A database $\db \in \dbs$, a parameter $\acc \in \R$, parameters $\priv, \privd, \accf \in [0,1]$, and the number of queries $\queries \in \mathbb{N}$.  Oracle access to $\update$, a $\maxupdates = \maxupdates(\acc)$-\idc for $\queryset$.}
\STATE{\textbf{Parameters:} $$\noisesize = \noisesize(\acc) := \frac{1000 \sqrt{\maxupdates(\acc)} \cdot \log(4/\delta)}{\eps} \qquad \threshold = \threshold(\acc) := 4 \noisesize(\acc) \cdot \log (2\queries/\accf).$$}
\STATE{Set $\dbstep{1} := \update(\emptyset, \cdot, \cdot)$, $\updates = 0$.}
\STATE{\textbf{For:} $t = 1, 2, \dots, \queries$}
\STATE{
\begin{enumerate}
\item Receive a query $\querystep{t} \in \queryset$ and compute $$\noisestep{t} \sim \Lap(\noisesize) \qquad \realstep{t} = \querystep{t}(\db) \qquad \noisyrealstep{t} = \querystep{t}(\db) + \noisestep{t} \qquad \fakestep{t} = \querystep{t}(\dbstep{t})$$
\item \textbf{If:} $|\noisyrealstep{t} - \fakestep{t}| \leq \threshold$ \textbf{then:} output $\fakestep{t}$ and set $\dbstep{t+1} = \dbstep{t}$ \\ \textbf{Else:} output $\noisyrealstep{t}$, set $\dbstep{t+1} = \update\left(\dbstep{t}, \querystep{t}, \noisyrealstep{t}\right)$, and set $\updates = \updates + 1$.
\item \textbf{If:} $\updates = \maxupdates(\acc)$ \textbf{then:} terminate.
\end{enumerate}}
\end{algorithmic}
\caption{Online Query Release Mechanism} \label{alg:onlineqr}
\end{algorithm}

\subsection{Privacy Analysis}
\begin{theorem}
Algorithm~\ref{alg:onlineqr} is $(\priv, \privd)$-differentially private.
\end{theorem}

\short{
\begin{proof}[Proof Sketch]
Our privacy analysis follows the approach of~\cite{HR10}.  Intuitively, we will try to classify the answers to the queries by the amount of ``information leaked about the database.''  This classification will lead to a bound on the total amount of information leaked, and a tighter bound can be deduced using Theorem~\ref{thm:eps-delta-composition}.

At a very high level, our argument can be thought of in two steps.  The first is to argue that the noise we add has large enough magnitude that the information leaked in the (small number of) ``update rounds'' is small.  This step is simple and follows from the bound on the number of update rounds and the well-known properties of the Laplace distribution.  The second step is to argue that the \emph{location} of the update rounds also leaks little information.  This second step is more difficult, and requires reasoning carefully about rounds that are ``close to update rounds.''

More specifically, though still informally we will consider three possible ranges for the value of the \emph{noise} $\noisestep{t}$ in each round $t = 1,2,\dots,\queries$..  Intuitively the three cases are as follows: 1) The noise is sufficiently small that there would never be an update, even if the input database were exchanged with an adjacent one.  Here we argue no privacy is leaked.  2) The noise is sufficiently large that there would always be an update, even if the database were exchanged.  In these rounds there is information leaked, but we also increment $\updates$, and thus cannot do too many of these before terminating.  3) The noise is intermediate, such that we do not do an update and increment $\updates$, but would if we switched to an adjacent database.  In principle there may be as many as $\queries$ such rounds, however it will turn out with high probability the number of such rounds is not much bigger than $\maxupdates$.

We then complete the proof by applying Theorem~\ref{thm:eps-delta-composition} to bound the \emph{total} privacy loss over the course of all the rounds.
\end{proof}
}

\full{
\begin{proof}
Our privacy analysis follows the approach of~\cite{HR10}.  Intuitively, we will consider each round of the mechanism individually, conditioned on the previous rounds and classify each round by the amount of ``information leaked'' from the database.  We will use this classification, as well as Azuma's Inequality to bound the total amount of information leaked.

Consider the vector
\begin{equation*}
\out = \set{\outstep{t}} =
\begin{cases} \noisyrealstep{t} & \text{if $t$ was an update round,}
\\
\bot &\text{otherwise.}
\end{cases}
\end{equation*}
Observe that $\out$ and the list of queries $\querystep{1}, \dots \querystep{k}$\footnote{We treat all the parameters of the mechanism, $\acc, \accf, \priv, \privd, k$ as well as the query sequence $\querystep{1}, \dots, \querystep{k}$ as public information.} are sufficient to reconstruct the internal state of the mechanism, and thus its output, in each round.  Therefore it will be sufficient to demonstrate that a mechanism that releases $\out$ is $(\priv, \privd)$-differentially private.

Fix any two adjacent databases $\db_1$ and $\db_2$, and let $V_1$ and $V_2$ denote the distributions on the vectors $\out$ when $\db_1$ and, $\db_2$ are the input database, respectively.  Also fix a vector $\out \in (\R \cup \bot)^\queries$.  We will use $\outsteps{t}$ to denote the first $t$ entries of the vector $\out$.  We will analyze the following \emph{privacy loss function} for each possible output vector $\out$
\begin{equation*}
\Psi(\out) = \log\left(\frac{V_1(\out)}{V_2(\out)}\right) = \sum_{t = 1}^{k} \log\left( \frac{V_1(\outstep{t} | \outsteps{t})}{V_2(\outstep{t} | \outsteps{t})} \right)
\end{equation*}

In each round $t = 1,2,\dots, \queries$, we define three ranges for the value of the \emph{noise} $\noisestep{t}$ that will describe whether or not we were ``never'', ``sometimes'', or ``always'' going to do an update in round $t$.  Specifically, let $\diffstep{t} = \querystep{t}(\db) - \querystep{t}(\dbstep{t-1})$.  Note that $\diffstep{t} = \realstep{t} - \fakestep{t}$ and that $\diffstep{t} + \noisestep{t} = \noisyrealstep{t} - \fakestep{t}$.  Now let
\begin{align*}
&\easystep{t} = (-\threshold - \diffstep{t} + \sigma, \threshold - \diffstep{t} - \sigma) \\
&\mediumstep{t} = [-\threshold - \diffstep{t} - \sigma, -\threshold - \diffstep{t} + \sigma] \cup [\threshold - \diffstep{t} - \sigma, \threshold - \diffstep{t}+\sigma] \\
&\hardstep{t} = (-\infty, -\threshold - \diffstep{t} - \sigma) \cup (\threshold - \diffstep{t} + \sigma, \infty)
\end{align*}
Intuitively, the event $\easystep{t}$ corresponds to values of the noise where $\noisyrealstep{t} -  \fakestep{t}$ is sufficiently small that switching databases could not cause an update.  In these rounds, $\outstep{t} = \bot$ with probability $1$ under both $V_1$ and $V_2$, so there is no privacy loss.  The event $\hardstep{t}$ corresponds to values of the noise where $\noisyrealstep{t} - \fakestep{t}$ is sufficiently large that switching databases could not prevent an update.  These rounds do leak information about the database, but the update will increment $\updates$, and thus there can only be $\maxupdates(\acc)$ such rounds.  The event $\mediumstep{t}$ are the problematic rounds.  In these rounds we may not update and increment $\updates$, thus in principle there may be an arbitrary number of these rounds.  However, $\noisyrealstep{t} - \fakestep{t}$ may be close enough to the update threshold that switching from $\db_1$ to $\db_2$ would cause an update.  Thus these rounds may incur privacy loss.  The remainder of the analysis relies on showing that there are not too many such rounds.

Now we make the following claims about the privacy loss in each type of round, based on the properties of the Laplace distribution and the way in which we defined the events $\easystep{t}, \mediumstep{t}, \hardstep{t}$.
\begin{claim}\label{clm:noloss}
For every $u \in \R \cup \bot$ and every $t =1,2,\dots,\queries$
$$
\log \left( \frac{V_1\left(\outstep{t} = u \mid \noisestep{t} \in \easystep{t}, \outsteps{t}\right)}{V_2\left(\outstep{t} = u \mid \noisestep{t} \in \easystep{t}, \outsteps{t}\right)} \right) = 0
$$
\end{claim}
\begin{proof}
Note that under both conditional measures, the probability of $\outstep{t} = \bot$ is $1$.
\end{proof}
\begin{claim}\label{clm:boundedloss}
For every $u \in \R \cup \bot$ and every $t = 1,2,\dots,\queries$
$$
\left| \log \left( \frac{V_1\left(\outstep{t} = u \mid \noisestep{t} \notin \easystep{t}, \outsteps{t}\right)}{V_2\left(\outstep{t} = u \mid \noisestep{t} \notin \easystep{t}, \outsteps{t}\right)} \right) \right| \leq 11\priv_0 = \frac{11\eps}{1000 \cdot \sqrt{\maxupdates} \cdot \log(4/\delta)}
$$
\end{claim}
The proof of this claim requires a straightforward analysis of the event $u = \bot$ under both conditional measures.  To not interrupt the flow of the larger proof, we defer the details until later.  The next claim states that the expected privacy loss is considerably smaller than the worst-case privacy loss.

\begin{claim}\label{clm:boundedexploss}
For every $u \in \R \cup \bot$ and every $t = 1,2,\dots,\queries$
$$
\Ex\left[\log \left( \frac{V_1\left(\outstep{t} = u \mid \noisestep{t} \notin \easystep{t}, \outsteps{t}\right)}{V_2\left(\outstep{t} = u \mid \noisestep{t} \notin \easystep{t}, \outsteps{t}\right)} \right) \right] \leq 242\priv_0^2 = \frac{121\priv^2}{500000 \cdot \maxupdates \cdot \log^2(4/\delta)}
$$
as long as $11\eps_0 \leq 1$.
\end{claim}
\begin{proof}
This claim follows from Claim~\ref{clm:boundedloss} and Theorem~\ref{thm:expectedprivacyloss}.
\end{proof}

In light of the previous claims, we want to bound the number of rounds in which $\easystep{t}$ does not occur.     Let $H = \card{t \mid \noisestep{t} \not\in \easystep{t}}$.
\begin{claim} \label{clm:overthreshold} For every $t =1,2,\dots, \queries$
$$
\prob{\noisestep{t} \in \hardstep{t} \mid \noisestep{t} \not\in \easystep{t}, \outsteps{t}} \geq 1/8.
$$
\end{claim}
\begin{proof}
\begin{align*}
&\prob{\noisestep{t} \in \hardstep{t} \mid \noisestep{t} \in \hardstep{t} \cup \mediumstep{t}, \outsteps{t}}
= \frac{\prob{\noisestep{t} \in \hardstep{t}, \outsteps{t}}}{\prob{\noisestep{t} \in \hardstep{t} \cup \mediumstep{t}, \outsteps{t}}} \\
&= \frac{\int_{T - R(t) + \sigma}^{\infty} \exp(-z / \noisesize) dz}{\int_{T - R(t) - \sigma}^{\infty} \exp(-z / \noisesize) dz}
= \frac{\exp(-(\threshold - \diffstep{t} + \sigma) / \noisesize)}{\exp(-(\threshold - \diffstep{t} - \sigma)/\noisesize)} \\
&= \exp(-2) \geq 1/8
\end{align*}
\end{proof}
\begin{claim}\label{clm:boundedhardrounds}
With probability $1-\delta/2$, $|H| \leq 16 \maxupdates \log(4/\delta)$.
\end{claim}
\begin{proof}
Claim~\ref{clm:overthreshold} implies that $\Ex[|H|] \leq 8
\maxupdates$.  Note that conditioned on the events of the previous rounds, the events $\noisestep{t} \in \hardstep{t}$ and $\noisestep{t} \in \mediumstep{t} \cup \hardstep{t}$ depend only on the coin tosses used to generate $\noisestep{t}$, which are independent of all of the other rounds.  Thus we can show that the random variable $|H|$ is dominated by a related random variable in which we do the following: In every round $t$ with $\noisestep{t} \in \mediumstep{t} \cup \hardstep{t}$, flip a coin $c^{(t)}$ such that $\prob{c^{(t)} = 1} = 1/8$.  Let $H'$ be defined identically to $H$ but in the process where we terminate the algorithm only when $\sum_{t=1}^{r} c^{(t)} = \maxupdates$, rather than the actual termination condition $\updates = \maxupdates$.  Since, by Lemma~\ref{clm:overthreshold}, we know that the probability $\noisestep{t} \in \hardstep{t}$ conditioned on $\noisestep{t} \in \mediumstep{t} \cup \hardstep{t}$ is at least $1/8$, we can couple these processes to ensure that $c^{(t)} = 1 \Longrightarrow \noisestep{t} \in \hardstep{t}$.  Thus, our new process will terminate no sooner than the actual algorithm for every choice of random coins, and $|H'|$ dominates $|H|$ in CDF.

Now it suffices to show that $|H'| \leq 2\Ex[|H'|]\log(4/\delta) \leq
16\maxupdates\log(4/\delta)$ with probability least $1-\delta/2$.
By a Chernoff
bound\footnote{A form of the Chernoff bound states that for independent
  $\{0,1\}$-random variables $X_1, \dots, X_n$, with $X = \sum_{i=1}^{n}
  X_i$ and $\mu = \Ex[X]$, $\prob{X < (1-\gamma)\mu} < \exp(-\mu \gamma^2 / 2)$.  From this we deduce that for $\mu \geq 2 \log(4/\gamma)$, $\prob{X < \mu/\log(1/\gamma)} < \gamma$.}, the probability that $\sum_{t \in H'} c^{(t)} \geq (1/ 16 \log(4/\delta)) |H'|$ is at least
$1 - \delta/2$.  Thus with probability at least $1-\delta/2$ we have
$|H'| \leq 16 \maxupdates \log(4/\delta)$.
\end{proof}

We now give a high-probability bound on the total privacy loss, conditioned on the event that $|H| \leq 16 \maxupdates \cdot \log(4/\delta)$.
\begin{claim}\label{clm:smalltotalloss}
If $|H| \leq 16 \maxupdates \cdot \log(4/\delta)$ then
$$\prob{|\Psi(\out)| > \priv} \leq \privd/2.$$
\end{claim}
\begin{proof}
The expected total privacy loss is
\begin{align}
\Ex\left[\Psi(\out)\right]
= {} &\Ex\left[\sum_{t = 1}^{k} \log\left( \frac{V_1(\outstep{t} | \outsteps{t})}{V_2(\outstep{t} | \outsteps{t})} \right) \right] = \sum_{t=1}^{k} \Ex\left[\log\left( \frac{V_1(\outstep{t} | \outsteps{t})}{V_2(\outstep{t} | \outsteps{t})} \right) \right] \notag \\
\leq {} &\sum_{t = 1}^{k} \prob{\easystep{t} \mid \outsteps{t}} \cdot \Ex\left[\log\left( \frac{V_1(\outstep{t} | \easystep{t}, \outsteps{t})}{V_2(\outstep{t} | \easystep{t}, \outsteps{t})} \right) \right] \notag \\
&+  \prob{\neg \easystep{t} \mid \outsteps{t}}  \cdot \Ex \left[  \log\left( \frac{V_1(\outstep{t} | \neg \easystep{t}, \outsteps{t})}{V_2(\outstep{t} | \neg \easystep{t}, \outsteps{t})} \right)\right] \label{concavity} \\
\leq {} &\sum_{t=1}^{k}  \prob{\neg \easystep{t} \mid \outsteps{t}}  \cdot 242 \priv_0^2 \notag \\
\leq {} &1936\maxupdates \priv_0^2 = \frac{191\priv^2}{62500 \cdot \log^2(4/\privd)} \leq \frac{\eps}{2} \notag
\end{align}
where~\eqref{concavity} follows from the convexity of relative entropy, and the final inequality follows from Claim~\ref{clm:overthreshold} and the fact that $\Ex[|H|] \leq 8$.

Conditioning on the coins of the mechanism we have
$$
\Psi(\out) = \sum_{t \in H} \log\left( \frac{V_1(\outstep{t} | \outsteps{t})}{V_2(\outstep{t} | \outsteps{t})} \right)
$$
and, by Claim~\ref{clm:boundedloss}, each term in the sum is at most $\priv_0$ in absolute value.  Thus we  can apply Azuma's Inequality\footnote{Azuma's Inequality states that for a sequence of random variables $X_0, X_1, \dots, X_n$, s.t. $|X_i - X_{i-1}| \leq \eta$ for $i = 1,2,\dots, n$, $\prob{|X_n - X_0| > \gamma} \leq 2\exp(-\gamma^2 / n\eta^2)$} to $\Psi(\out)$ to show
\begin{align*}
\prob{|\Psi(\out)| > \priv}
\leq {} &\prob{ |\Psi(\out) - \Ex\left[\Psi(\out)\right]| > \priv/2} \\
\leq {} &2 \exp\left( - \frac{\priv^2}{2 |H| \priv_0^2}\right) \\
\end{align*}
If we condition on the event that $|H| \leq 16 \maxupdates \log(4/\privd)$ then we have
$$
\prob{|\Psi(\out)| > \priv} \leq 2 \exp\left( - \log(4/\delta) \right) \leq \privd/2,
$$
which proves the claim.
\end{proof}
Claims~\ref{clm:boundedhardrounds} and~\ref{clm:smalltotalloss} suffice to prove the Theorem.
\end{proof}

We now give a proof of Claim~\ref{clm:boundedloss}.
\begin{claim}[Claim~\ref{clm:boundedloss}, restated]
For every $u \in \R \cup \bot$ and every $t = 1,2,\dots,\queries$
$$
\left| \log \left( \frac{V_1\left(\outstep{t} = u \mid \noisestep{t} \notin \easystep{t}, \outsteps{t}\right)}{V_2\left(\outstep{t} = u \mid \noisestep{t} \notin \easystep{t}, \outsteps{t}\right)} \right) \right| \leq 11\priv_0 = \frac{11\eps}{1000 \cdot \sqrt{\maxupdates} \cdot \log(4/\delta)}
$$
\end{claim}
\begin{proof}[Proof of Claim~\ref{clm:boundedloss}]
First we will bound the left-hand-side in the case of $u \in \R$.
\begin{align}
\left| \log \left( \frac{V_1\left(\outstep{t} = u \mid \noisestep{t} \notin \easystep{t}, \outsteps{t}\right)}{V_2\left(\outstep{t} = u \mid \noisestep{t} \notin \easystep{t}, \outsteps{t}\right)} \right) \right|
&= \left| \log \left( \frac{\exp( -| u - \querystep{t}(\db_1) | / \noisesize )}{\exp( - |u - \querystep{t}(\db_2)| / \noisesize)} \right) \right| \notag \\
&\leq \log \left( \exp(1/\noisesize) \right)= 1/\noisesize \label{eq:sens} \\
&=\eps_0 \notag
\end{align}
where inequality (\ref{eq:sens}) follows because the sensitivity of $\querystep{t}$ is
bounded above by~$1$.  Now we will consider the case of $u = \bot$.

\begin{eqnarray*}
&& \left| \log \left( \frac{V_1\left(\outstep{t} = u \mid \noisestep{t} \notin \easystep{t}, \outsteps{t}\right)}{V_2\left(\outstep{t} = u \mid \noisestep{t} \notin \easystep{t}, \outsteps{t}\right)} \right) \right| \\
&=& \left| \log \left( \frac{\int_{-R(t)-T}^{-R(t)-T+\sigma}\exp(-|u|/\sigma) du + \int_{-R(t)+T-\sigma}^{-R(t)+T}\exp(-|u|/\sigma)du}
{\int_{-R(t)-T+1}^{-R(t)-T+\sigma}\exp(-|u|/\sigma)du + \int_{-R(t)+T-\sigma}^{-R(t)+T-1}\exp(-|u|/\sigma)du} \right) \right|\\
&\leq& \left| \log \left( \frac{\int_{-R(t)-T}^{-R(t)-T+\sigma}\exp(-|u|/\sigma) du}{\int_{-R(t)-T+1}^{-R(t)-T+\sigma}\exp(-|u|/\sigma)du} +
\frac{\int_{-R(t)+T-\sigma}^{-R(t)+T}\exp(-|u|/\sigma)du}{\int_{-R(t)+T-\sigma}^{-R(t)+T-1}\exp(-|u|/\sigma)du}\right) \right| \\
&=&\left| \log \left( 2 + \frac{\int_{-R(t)-T}^{-R(t)-T+1}\exp(-|u|/\sigma)du}{\int_{-R(t)-T+1}^{-R(t)-T+\sigma}\exp(-|u|/\sigma)du}
+\frac{\int_{-R(t)+T-1}^{-R(t)+T}\exp(-|u|/\sigma)du}{\int_{-R(t)+T-\sigma}^{-R(t)+T-1}\exp(-|u|/\sigma)du}
\right) \right|\\
&\leq& \left| \log \left( 2\left(1+\frac{e}{\sigma-1}\right)\right) \right| \\
&\leq& \frac{2e}{\sigma-1} \leq \frac{4e}{\sigma} \leq 11\priv_0
\end{eqnarray*}
\end{proof}
}

\subsection{Utility Analysis}
\begin{theorem} Let $\db \in \dbs$ be any database. And $\update$ be a
  $\maxupdates(\acc)$-\idc for query class $\queryset$. Then for any
  $\beta, \priv, \privd > 0$, Algorithm~\ref{alg:onlineqr} is
  $\left(\frac{5\threshold(\acc)}{4},\beta\right)$-accurate for
  $\queryset$, as long as $\threshold(\acc) \in [4\acc/3, 2\acc]$.
\end{theorem}

\short{
\begin{proof}[Proof sketch]
Roughly, the argument is as follows:  Assume we did not add any noise to the queries.  Then we would answer each query with the exactly-correct answer $\realstep{t}$ or with $\fakestep{t}$ so long as $\fakestep{t}$ is sufficiently close to $\realstep{t}$.  Essentially, all we do in the proof is show that this intuition remains correct when noise is added.

When adding noise we answer with either $\realstep{t} + \noisestep{t}$ or $\fakestep{t}$, so long as $\fakestep{t}$ is sufficiently close to $\realstep{t} + \noisestep{t}$.  It is not hard to argue that $\noisestep{t}$ remains small in every round, and thus the answers in the latter case are not much less accurate than the answers in the former case.

What remains to be shown is that the mechanism does not terminate early due to the condition $\updates = \maxupdates$.  In order to do this, we show that the sequence of updates forms a \dus, and thus cannot be too long if $\update$ is an efficient \idc.  In order to do this, we argue that $\noisestep{t}$ is sufficiently small that the condition for performing an update ($|\realstep{t} + \noisestep{t} - \fakestep{t}| \geq \threshold$) is sufficient to ensure that the query is a good distinguisher ($|\realstep{t} - \fakestep{t}| \geq \acc$).
\end{proof}

In order to get the best accuracy parameters, one can just solve for the
equation $\acc = 3\threshold(\acc)/4$; substituting for $T(\cdot)$,
this is the same as solving the following equation for $\acc$:
$\acc = \frac{96 \sqrt{\maxupdates(\acc)} \log(4/\delta)
    \log(\queries/\beta)}{\eps}$.
Using this method we obtain bounds on the error for various IDCs, which are summarized both in Table~\ref{tab:results} and in the full version.
}

\full{
\begin{proof}
Roughly, the argument is as follows:  Assume we did not add any noise to the queries.  Then we would answer each query with the true answer $\realstep{t}$ or with $\fakestep{t}$ if $\fakestep{t}$ is sufficiently close to $\realstep{t}$.  Thus the only reason the mechanism would fail to be accurate is if it performs too many updates and has to terminate due to the condition $\updates = \maxupdates$.  But since we only invoke $\update$ when we find a query such that $|\querystep{t}(\db) - \querystep{t}(\dbstep{t}) |$ is large, we are actually generating a \dus, which cannot be too long if $\update$ is an efficient \idc.  To formalize this intuition we have to consider the effect of the noise on this process and show that with high probability the noise remains in a small enough range that this intuition is indeed correct.

Fix any $\acc, \threshold(\acc)$, such that $\threshold(\acc) \in
[4\acc/3, 2\acc]$. For brevity, we use $\threshold$ to denote
$\threshold(\acc)$.  First, we observe that, with probability
$1-\accf$, $$\max_{t=1,2,\dots,\queries} |\noisestep{t}| \leq
\threshold/4.$$ Indeed, by a direct calculation:
\begin{align*}
\prob{\max_{t = 1,2,\dots \queries} |\noisestep{t}| > \threshold/4}
&\leq \queries \cdot \prob{|\noisestep{1}| > \threshold/4} \\
&\leq 2\queries \exp\left( -\threshold/4\noisesize \right) \\
&= 2\queries \exp\left( - \log(2\queries/\accf) \right) \leq \beta
\end{align*}
For the rest of the proof we will condition on this event and show that for every $t = 1,2,\dots,\queries$,
$$
\left| \querystep{t}(\dbstep{t}) - \querystep{t}(\db) \right| \leq 5\threshold/4
$$

Assuming the algorithm has not yet terminated, in step 3 we answer each query with either $\fakestep{t}$ s.t.
\begin{align*}
\threshold
&\geq |\noisyrealstep{t} - \fakestep{t}| = |\querystep{t}(\db) + \noisestep{t} - \fakestep{t}| \\
&\geq |\querystep{t}(\db) - \fakestep{t}| - |\noisestep{t}| \geq |\querystep{t}(\db) - \fakestep{t}| - \threshold/4,
\end{align*}
(in which case the error is at most $5T/4$); or else we answer directly with
$\noisyrealstep{t}$, in which case
$$|\querystep{t}(\db) - \noisyrealstep{t}| = |\noisestep{t}| \leq
\threshold/4 \leq 5T/4.$$

Now it suffices to show that Algorithm~\ref{alg:onlineqr} does not
prematurely terminate (due to the condition $\updates = \maxupdates$)
before answering every query, and in particular that the sequence of
invocations of $\update$ form an $(\update, \db, \queryset, \acc,
\updates)$-\dus.  Indeed, if this were the case, then we'd be assured
(by Definition~\ref{def:idc}) that after $\maxupdates$ invocations of
$\update$, the resulting database $\db^*$ would be $(\acc,
\queryset)$-accurate. So in every subsequent round we'd have
$$|\noisyrealstep{t} - \fakestep{t}| = |\querystep{t}(\db) +
\noisestep{t} - \querystep{t}(\dbstep{t})| \leq |\querystep{t}(\db) -
\querystep{t}(\dbstep{t})| + |\noisestep{t}| \leq \acc + \threshold/4
\leq \threshold,$$
and we'd never make the $\maxupdates+1^{st}$ update.  So to complete the proof, we
show that we satisfy the properties in Definition~\ref{def:dus}. Firstly,
in every round in which we invoke $\update$,
\begin{align*}
\threshold < |\noisyrealstep{t} - \fakestep{t}| \leq |\querystep{t}(\db) - \querystep{t}(\dbstep{t})| + |\noisestep{t}| \leq |\querystep{t}(\db) - \querystep{t}(\dbstep{t})| + \threshold/4 \\
\Longrightarrow |\querystep{t}(\db) - \querystep{t}(\dbstep{t})| > 3T/4 \geq \acc
\end{align*}
so that the update sequence satisfies property 2 of Definition~\ref{def:dus}.
Secondly, we have already seen that in every round
$$|\noisyrealstep{t} - \querystep{t}(\db)| \leq \threshold/4 \leq \acc/2,$$ so that the update sequence satisfies property 3 of Definition~\ref{def:dus}.  Properties 1 and 4 of Definition~\ref{def:dus} follow by the construction of Algorithm~\ref{alg:onlineqr}.  This completes the proof.
\end{proof}

In order to get the best accuracy parameters, one can just solve for the
equation $\acc = 3\threshold(\acc)/4$; substituting for $T(\cdot)$,
this is the same as solving the following equation for $\acc$:
\begin{gather}
  \acc = \frac{3000 \sqrt{\maxupdates(\acc)} \log(4/\delta)
    \log(\queries/\beta)}{\eps}.
\end{gather}

\begin{corollary}
The Multiplicative Weights mechanism is $(\epsilon,\delta)$-differentially private and $(\acc,\beta)$ accurate for:
$$\acc = O\left(\frac{\sqrt{n}(\log |\univ|)^{1/4}\sqrt{\log(4/\delta)\log(k/\beta)}}{\sqrt{\epsilon}}\right)$$
The Median Mechanism is $(\epsilon,\delta)$-differentially private and $(\acc,\beta)$ accurate for:
$$\acc = O\left(\frac{\sqrt{n}(\log |\univ|\log k)^{1/4}\sqrt{\log (4/\delta)\log(k/\beta)}}{\sqrt{\epsilon}}\right)$$
\end{corollary}
\begin{proof}
The multiplicative weights and median mechanism subroutines are given in Appendix \ref{app:other-IDCs}. By Theorem \ref{thm:MW-Decomposition}, the multiplicative weights subroutine is a $\maxupdates(\acc)$-IDC for $\maxupdates(\acc) = 4n^2\log|\univ|/\acc^2$. By Theorem \ref{thm:MM-Decomposition}, the median mechanism subroutine is a $\maxupdates(\acc)$-IDC for $\maxupdates(\acc) = n^2\log k\log |\univ|/\acc^2$. The bounds then follow simply by solving for $\acc$ in the expression $ \acc = \frac{3000 \sqrt{\maxupdates(\acc)} \log(4/\delta)\log(\queries/\beta)}{\eps}$.
\end{proof}
\begin{remark}
We note that for the setting in which the database represents the edge
set of a graph $G = (V,E)$, and the class of queries we are interested
in is the set of all cut queries, this corresponds to an error bound of  $\tilde{O}(\sqrt{|E||V|}\log(V)^{1/4}/\sqrt{\epsilon})$.
\end{remark}
}

\section{An \IDC~Based on Frieze/Kannan}
In this section we describe and analyze an \idc based on the Frieze/Kannan ``cut decomposition''~\cite{FK99}.  Although the style of analysis we use was originally applied specifically to cuts in~\cite{FK99}, we use a generalization of their argument to arbitrary linear queries.  To our knowledge, such a generalization was first observed in~\cite{TTV09}.

\begin{algorithm}
\noindent\textbf{$\update^{FK}_{\acc}(\db, \query, \noisyreal)$:}
\begin{algorithmic}
\STATE{\textbf{If:} $\db = \emptyset$ \textbf{then:} output $\db' = \emptyset$}
\STATE{\textbf{Else if:} $\query(\db) - \noisyreal > 0$ \textbf{then:} output $\db' = \db - \frac{\acc}{|\univ|} \cdot \query$}
\STATE{\textbf{Else if:} $\query(\db) - \noisyreal < 0$ \textbf{then:} output $\db' = \db + \frac{\acc}{|\univ|} \cdot \query$}
\end{algorithmic}
\caption{The Frieze/Kannan-based IDC} \label{alg:fk}
\end{algorithm}
Note that the sum in Algorithm~\ref{alg:fk} denotes vector addition.

\begin{theorem}
Let $\db \in \dbs$ be a dataset.  For any $\acc > 0$, $\update^{FK}_{\acc}$ is a $\maxupdates(\acc)$-\idc for a class of linear queries $\queryset$, where $\maxupdates(\acc)= \frac{ \|\db\|_2^2 |\univ|}{\acc^2}$.
\end{theorem}

\short{
\begin{proof} [Proof sketch]
Let $\db \in \dbs$ be any database and let $\set{(\dbstep{t}, \querystep{t}, \noisyrealstep{t})}_{t=1,\dots, \updates}$ be $(\update^{FK}_{\acc}, \db, \queryset, \acc, \maxupdates)$-\dus (Definition~\ref{def:dus}).  We want to show that $\updates \leq \|\db\|_2^2 |\univ| /\acc^2$.  Specifically, that after $\|\db\|_2^2 |\univ| /\acc^2$ invocations of $\update^{FK}_{\acc}$, the database $\dbstep{\|\db\|_2^2 |\univ|/\acc^2}$ is $(\acc, \queryset)$-accurate for $\db$, and thus there cannot be a sequence of longer than $\|\db\|_2^2 |\univ|/\acc^2$ queries that satisfy property 2 of Definition~\ref{def:dus}.

In order to formalize this intuition, we use a potential argument as in~\cite{FK99} to show that for every $t = 1,2,\dots,\maxupdates$, $\dbstep{t+1}$ is significantly closer to $\db$ than $\dbstep{t}$.  Specifically, our potential function is the $L_2^2$ norm of the database $\db - \dbstep{t}$, defined as
$\|\db\|_2^2 = \sum_{i \in \univ} \db(i)^2.$
Observe that $\|\db - \dbstep{1}\|_2^2 = \|\db\|_2^2 $, and
$\|\db\|_2^2 \geq 0$.  Thus it will suffices to show, as we do in the full proof, that in every
step, the potential decreases by $\acc^2 / |\univ|$.
\end{proof}
}

\full{
\begin{proof}
Let $\db \in \dbs$ be any database and let $$\set{(\dbstep{t}, \querystep{t}, \noisyrealstep{t})}_{t=1,\dots, \updates}$$ be $(\update^{FK}_{\acc}, \db, \queryset, \acc, \maxupdates)$-\dus (Definition~\ref{def:dus}).  We want to show that $\updates \leq \|\db\|_2^2 |\univ| /\acc^2$.  Specifically, that after $\|\db\|_2^2 |\univ| /\acc^2$ invocations of $\update^{FK}_{\acc}$, the database $\dbstep{\|\db\|_2^2 |\univ|/\acc^2}$ is $(\acc, \queryset)$-accurate for $\db$, and thus there cannot be a sequence of longer than $\|\db\|_2^2 |\univ|/\acc^2$ queries that satisfy property 2 of Definition~\ref{def:dus}.

In order to formalize this intuition, we use a potential argument as in~\cite{FK99} to show that for every $t = 1,2,\dots,\maxupdates$, $\dbstep{t+1}$ is significantly closer to $\db$ than $\dbstep{t}$.  Specifically, our potential function is the $L_2^2$ norm of the database $\db - \dbstep{t}$, defined as
$$\|\db\|_2^2 = \sum_{i \in \univ} \db(i)^2.$$
Observe that $\|\db - \dbstep{1}\|_2^2 = \|\db\|_2^2 $, and
$\|\db\|_2^2 \geq 0$.  Thus it suffices to show that in every
step, the potential decreases by $\acc^2 / |\univ|$.  We analyze the
case where $|\querystep{t}(\dbstep{t})| < \noisyrealstep{t}$, the
analysis is the opposite case will be similar.  Let $\diffstep{t} =
\dbstep{t} - \db$.  Observe that in this case we have
$$|\querystep{t}(\diffstep{t})| \geq \acc$$
and
$$\querystep{t}(\diffstep{t}) \geq \noisyrealstep{t} - \querystep{t}(\dbstep{t}) - \acc/2 > -\acc/2.$$
Thus we must have
$$\querystep{t}(\diffstep{t}) \geq \acc.$$
Now we can analyze the drop in potential.
\begin{align}
\|\diffstep{t}\|_2^2 - \|\diffstep{t+1}\|_2^2
&= \|\diffstep{t}\|_2^2 - \|\diffstep{t} - (\acc/|\univ|) \cdot \querystep{t} \|_2^2 \notag \\
&= \sum_{i \in \univ} \diffstep{t-1}(i)^2 - \left(\diffstep{t}(i,j) -  (\acc/|\univ|)\cdot \querystep{t}(i) \right)^2 \notag \\
&= \sum_{i \in \univ} \left( \frac{2\acc}{|\univ|} \cdot \diffstep{t}(i) \querystep{t}(i) - \frac{\acc^2}{|\univ|^2} \querystep{t}(i)^2 \right) \notag \\
&= \frac{2\acc}{|\univ|} \querystep{t}(\diffstep{t}) - \frac{\acc^2}{|\univ|^2} \sum_{i \in \univ} \querystep{t}(i)^2 \notag \\
&\geq \frac{2\acc}{|\univ|} \querystep{t}(\diffstep{t}) - \frac{\acc^2}{|\univ|^2} |\univ| \notag \\
&\geq \frac{2 \acc^2}{|\univ|} - \frac{\acc^2 }{|\univ|} = \frac{\acc^2}{|\univ|} \notag
\end{align}
This bounds the number of steps by $\|\db\|_2^2 |\univ| /\acc^2$, and completes the proof.
\end{proof}
}

\begin{corollary} Algorithm~\ref{alg:onlineqr}, instantiated with
  $\update^{FK}_{\gamma}$ for $\gamma = O\left(\eps^{-1/2}n_2^{1/4}
    |\univ|^{1/4} \sqrt{\log (\queries/\beta)}\right)$ is $(\priv,
  \privd)$-differentially private and an $(\acc, \accf)$-accurate
  interactive release mechanism for query set $\queryset$ with
$\acc = O\left( \frac{n_2^{1/4} |\univ|^{1/4} \sqrt{\log(k/\beta) \log(1/\delta)}}{\sqrt{\eps}} \right)$
where $n_2 = \|\db\|_2^2$.
Note that for databases that are subsets of the data universe (rather than multisets), $n_2 = n$.
\end{corollary}

\begin{remark}
We note that for the setting in which the database represents the edge set of a graph $G = (V,E)$, and the class of queries we are interested in is the set of all cut queries, this bounds corresponds to $\tilde{O}(|V||E|^{1/4}/\sqrt{\epsilon})$. This is an improvement on the bound given by the multiplicative weights IDC for dense graphs: when $|E| \geq \Omega(|V|^2/\log|V|)$.
\end{remark}


%% file: randomizedresponse.tex
\section{Results for Synthetic Data}
\label{sec:synthetic-data}

In this section, we consider the more demanding task of efficiently
releasing \emph{synthetic data} for the class of cut queries on graphs.
\full{The task at hand here is to actually generate another graph that
approximates the private graph with respect to cuts. Such a graph can
then simply be released to data analysts, who can examine it at their
leisure. This is preferable to the interactive setting, in which an
actual graph is never produced, and a central stateful API must be
maintained to handle queries as they come in from data analysts.} Our
algorithm is simple, and is based on releasing a noisy histogram. Note
that for a graph, $|\univ| = {|V| \choose 2}$, and $\db = E$, so as long
as $|E| = \Omega(|V|)$, the universe is at most a polynomial in the
database size. (Moreover, it is easy to show that there does not exist
any $(\epsilon,0)$-private mechanism that has error $o(|V|)$, so the
only interesting cases are when $|E| = \Omega(|V|)$.)

Consider a database whose elements are drawn from $\univ$; we represent
this as a vector (histogram) $\db \in \mathbb{N}^{|\univ|}$. Let
$\widehat{\db} = \db + (Y_1,\ldots,Y_{|\univ|})$ be a ``noisy''
database, where each $Y_i \sim \textrm{Lap}(1/\varepsilon)$ is an
independent draw from the Laplace distribution. Note that by Theorem
\ref{thm:laplace-privacy}, the procedure which on input $\db$ releases
the noisy database $\widehat{\db}$ preserves $(\epsilon,0)$-differential
privacy. This follows because the histogram vector can be viewed as simply the evaluation of the identity query $\query:\dbs\rightarrow\dbs$, which can be easily seen to be 1-sensitive. At this stage, we could release $\widehat{\db}$ and be
satisfied that we have designed a private algorithm. There are two
issues: first, we must analyze the utility guarantees that
$\widehat{\db}$ has with respect to our query set $\queryset$. Second,
$\widehat{\db}$ is not quite synthetic data. It will be a vector with
possibly negative entries, and so does not represent a
histogram. Interpreted as a graph, it will be a weighted graph with
negative edge weights. This may not be satisfactory, so we must do a
little more work.

The utility guarantee of this procedure over the collections $\queryset$
of linear queries is also not difficult; i.e., each query $\query \in
\queryset$ is a vector in $[0,1]^{|\univ|}$, and on any database $\db$
evaluates to $\query(\db) = \langle \query, \db\rangle$.

\begin{lemma}
  \label{thm:rr-laplace}
  Suppose that $\queryset \subseteq [0,1]^{|\univ|}$ is some collection
  of linear queries. For the case $|\queryset| \leq (\beta/2)\,
  2^{|\univ|/6}$, it holds that with probability at least $1- \beta$,
  \[ | \query(\db) - \query(\widehat{\db}) | \leq \varepsilon^{-1}
  \sqrt{6|\univ| \log (|\queryset|/\beta)} \] for every query $\query \in
  \queryset$. For general $\queryset$, the error bound is
  $O(\varepsilon^{-1} \sqrt{|\univ| \log (|\univ|/\beta) \log
    (|\queryset|/\beta)})$.
\end{lemma}
\short{The proof of this lemma uses standard techniques and is deferred to the full version.}
\full{
\begin{proof}
  Note that $(\query(\widehat{\db}) - \query(\db)) = \langle \query, \db
  - \widehat{\db} \rangle \sim \sum_i q_i\, Y_i$, where each random
  variable $Y_i \sim \Lap(1/\varepsilon)$, and $q_i \in [0,1]$ is the $i^{th}$ coordinate of query
  $\query$. By a tail bound for sums of Laplace random variables
  (Theorem~\ref{thm:conc}), we know that
  \[ \textstyle \Pr[ |\sum_{i = 1}^{|\univ|} q_iY_i| \geq \acc ] \leq
  2\exp(- \acc^2\varepsilon^2/6|\univ|), \] as long as $\acc \leq
  |\univ|/\varepsilon$. If we set $\acc = \varepsilon^{-1}
  \sqrt{12|\univ| \log (2|\queryset|/\beta)}$ the probability bound on the
  right hand side at most $\beta$, and the condition $\acc \leq
  |\univ|/\varepsilon$ translates to $|\queryset| \leq (\beta/2)\,
  2^{|\univ|/6}$.

  The proof for the general case, where we do not assume a bound on the
  size $|\queryset|$, loses an extra factor of
  $O(\sqrt{\log(|\univ|/\beta)})$. Indeed, with probability at least
  $1-\beta/2$, each of the absolute values $|Y_i|$'s are at most $L =
  O(1/\varepsilon) \log (|\univ|/\beta)$. Now, conditioning on this
  event happening, the sum $\sum_{i = 1}^{|\univ|} q_i Y_i$ behaves like
  a sum of $|\univ|$-many independent $[-L,L]$-bounded random variables
  with mean $0$: in this case, $\Pr[ |\sum_i q_iY_i| > \acc] \leq
  2e^{- \Omega(\acc^2/(L^2|\univ|))} = e^{-\Omega(\acc^2
    \varepsilon^2/(|\univ| \log (|\univ|/\beta)))}$ by a standard
  Chernoff bound. Now setting $\acc = O(\varepsilon^{-1} \sqrt{|\univ|
    \log (|\univ|/\beta) \log (|\queryset|/\beta)})$ causes this
  probability to be at most $\beta/2$; by a union bound, the probability
  of large deviations is at most $\beta$.
\end{proof}
}

In summary, note that the bounds on the error are $\approx
\varepsilon^{-1} \sqrt{|\univ| \log |\queryset|}$, with some correction
terms depending on whether the size of the query set is at most
$2^{O(|\univ|)}$ or larger.

\subsection{Randomized Response and Synthetic Data for Cut Queries}

For the case of cuts in graph on a vertex set $V$, the database is a
vector in $\smash{\{0,1\}^{\binom{|V|}{2}}}$, and the noisy database
just adds independent $\Lap(1/\varepsilon)$ noise to each bit
value. Since the query set $\queryset_{cuts}$ has size $2^{2|V|}$,
(namely it consists of all $(S,T)$ pairs), we have $|\queryset_{cuts}| \ll (\beta/2)
2^{|\univ|/6}$ for all reasonable $\beta$ and $|V|$, we can use the
randomized response analysis above to get accuracy
\[ \textstyle O\left(\left(\binom{|V|}{2} \log
  (|\queryset_{cuts}|/\beta)\right)^{1/2}/\varepsilon\right) = O((|V|^{3/2} +
|V|\log 1/\beta)/\varepsilon) \] with probability at least $1-\beta$.
In fact, one can give a slightly tighter analysis where the accuracy
depends on the size of the sets $S,T$---by observing that the number of
random variables participating in a cut query $(S,T)$ is exactly
$|S||T|$, one can show that the accuracy for all cuts is whp
$O(\varepsilon^{-1} \sqrt{|V||S||T|})$.

Viewing the noisy database $\widehat{\db}$ as a weighted graph
$\widehat{G}$, where the weight of $(u,v)$ is $\mathbf{1}_{(u,v) \in
  E(G)} + \Lap(1/\varepsilon)$, note that $\widehat{G}$ has negative
weight edges and hence cannot be considered synthetic data. We can
remedy the situation (using the idea of solving a suitable linear
program~\cite{BCDKMT07,DNRRV09}):

\begin{lemma}[\textbf{Synthetic Data for Cuts}]
  \label{lem:rr-LP}
  There is a computationally efficient $(\varepsilon,0)$-differentially
  private randomized algorithm that takes a unweighted graph $G$ and
  outputs a synthetic graph $G'$ such that, with high
  probability, $\|G - G'\|_C \leq O(|V|^{3/2}/\varepsilon)$---all cuts
  in $G$ and $G'$ are within $O(|V|^{3/2}/\varepsilon)$ additive error.
\end{lemma}
\short{The proof is deferred to the full version, but the idea is straightforward: we write a linear program with exponentially many constraints to solve for a synthetic database, and use an SDP of Alon and Naor for the cut-norm problem as a separation oracle to solve the LP.}
\full{
\begin{proof}
First we construct the noisy datastructure $\widehat{D}$ by perturbing each entry of $\db$ with independent noise drawn from Lap$(1/\epsilon)$. All further operations will be conducted on $\widehat{D}$, and so the entire algorithm will be $(\epsilon,0)$-differentially private. let $z_{i,j}$ denote the $i,j$'th entry of $\widehat{D}$: i.e. $z_{i,j} = \db[i,j] + \textrm{Lap}(1/\epsilon)$.
 Let us condition on the event that for
  every cut $(S,T)$, the additive error bound is
  $O(|V|^{3/2}/\varepsilon)$. Now define the following LP:
  \begin{align*}
    &\min \lambda \\
    \text{such that} \sum_{ij \in S \times T} (x_{ij} - z_{ij}) &\leq
    \lambda  \qquad \qquad \forall S,T \\
    \sum_{ij \in S \times T} (z_{ij} - x_{ij}) &\leq
    \lambda  \qquad \qquad \forall S,T \\
    x_{ij} &\in [0,1].
  \end{align*}
  There exists a feasible solution to this LP with $\lambda =
  O(|V|^{3/2}/\varepsilon)$, since we can just use the original graph to
  get the solution $x_{ij} = \mathbf{1}_{(ij \in E(G))}$. Now if we
  solve the LP, and output the optimal feasible solution to the LP, it
  would be a weighted graph $G'$ such that $\| G - G' \|_C \leq
  O(|V|^{3/2}/\varepsilon)$.

  Since the LP has exponentially many constraints, it remains to show
  how to solve the LP. Define the matrix $A$ with $A_{ij} = x_{ij} -
  z_{ij}$, and define $A(S,T) := \sum_{i \in S, j \in T} A_{ij}$, then
  the separation oracle must find sets $S, T$ such that $| A(S,T) |$ is
  larger than $\lambda$. Equivalently, it suffices 
  to approximately
  compute the cut norm of the matrix $A$.  There is a constant-factor
  approximation algorithm of Alon and Naor for the cut norm
  problem~\cite{AN03}; using this we can solve the LP above to within
  constant factors of optimum.
\end{proof}
\begin{remark}
The procedure outlined above results in outputting a weighted graph (with non-negative edge weights) $\db' \in [0,1]^{|V|\times|V|}$. Note that if the original graph was unweighted: $\db \in \{0,1\}^{|V|\times|V|}$ and it is desired to output another unweighted graph, we can simply randomly round $\db'$ to an integral solution in the obvious way. This does not incur any asymptotic loss in the stated accuracy bound.
\end{remark}
}

\full{
\subsection{A Spectral Solution, and Rank-$1$ Queries}
The Alon-Naor algorithm involves solving SDPs which are computationally
intensive, but we can avoid that by using the tighter accuracy bound of
$O(\varepsilon^{-1} \sqrt{|V||S||T|})$ we proved. Consider the modified
LP:
\begin{align*}
  &\min \lambda \\
  \text{such that} \sum_{ij \in S \times T} (x_{ij} - z_{ij}) &\leq
  \lambda \sqrt{|S||T|} \qquad \qquad \forall S,T \\
  \sum_{ij \in S \times T} (z_{ij} - x_{ij}) &\leq
  \lambda \sqrt{|S||T|} \qquad \qquad \forall S,T \\
  x_{ij} &\in [0,1].
\end{align*}
Again, this LP has a feasible solution (whp) with $\lambda =
O(|V|^{1/2}/\varepsilon)$. And solving the LP to within a factor of
$\rho$, and outputting a near-optimal feasible solution to the LP would
give a synthetic weighted graph $G'$ such that $\| G - G' \|_C \leq
O(\rho\cdot \sqrt{|V||S||T|}/\varepsilon) =
O(\rho\cdot|V|^{3/2}/\varepsilon)$. To this end, define the
\emph{normalized cut norm} as
\[ \| A\|_{NC} := \max_{S, T} \frac{|A(S,T)|}{\sqrt{|S||T|}}; \] Now the
separation problem is to find $S,T$ approximately maximizing the
normalized cut norm. For this we use a theorem of Nikiforov~\cite{Niki}
which says that if $\sigma_1(A)$ is the top singular value of $A$ (and
$\|A\|_2$ is $A$'s spectral norm) then
\begin{gather}
  \| A\|_{NC} \leq \sigma_1(A) = \| A \|_2 \leq \| A\|_{NC} \cdot O(\log
  |V|).\label{eq:1}
\end{gather}
There is also a polynomial-time algorithm that given the top singular
value/vector for $A$, returns a normalized cut $(S',T')$ of value
$|A(S', T')|/\sqrt{|S'||T'|} \geq \sigma_1(A) / O(\log |V|) \geq
\|A\|_{NC}/O(\log |V|)$. Using this as a separation oracle we can solve
the LP to within $\rho = O(\log |V|)$ of the optimum, and hence get an
additive error of $O(\varepsilon^{-1} \log |V| \sqrt{|V||S||T|} ) =
O(\varepsilon^{-1} |V|^{3/2} \log |V|)$.

\begin{remark}
We note that the theorem of Nikiforov quoted above \cite{Niki} also implies that the synthetic graph released by our algorithm is useful for the (infinite) set of rank-1 queries $\queryset_{r1}$ as well as the set of cut queries, with only an $O(\log|V|)$ factor loss in the additive approximation for each query.
\end{remark}
}
\full{
\subsection{A Tail Bound for Laplace Distributions}
\label{sec:tail-bound}

The following tail bound for Laplace random variables uses standard
techniques, we give it here for completeness.
\begin{theorem}
  \label{thm:conc}
  Suppose $\{Y_i\}_{i = 1}^k$ are i.i.d.\ $\Lap(b)$ random variables,
  and scalars $q_i \in [0,1]$. Define $Y := \sum_i q_iY_i$. Then
  \begin{gather}
    \Pr[ Y \geq \acc ] \leq \begin{cases}
      \exp\left( - \frac{\acc^2}{6kb^2}
      \right).    &\qquad\text{if $\acc \leq kb$} \\
      \exp\left( - \frac{\acc}{6b}
      \right).    &\qquad\text{if $\acc > kb$}
    \end{cases}
  \end{gather}
\end{theorem}
\begin{proof}
  Suppose $Y_1, Y_2, \ldots, Y_k$ are i.i.d. Laplace$(b)$ random
  variables, and $q_1, q_2, \ldots, q_k$'s are scalars in $[0,1]$. We
  now give a tail bound for $Y := \sum_i q_iY_i$. It is useful to recall that the moment
  generating function for a Laplace random variable $Y \sim
  \Lap(b)$ is $E[e^{tY}] = 1/(1 - b^2t^2)$ for $|t| < 1/b$, and also that if $Y \sim
  \Lap(b)$, then $cY \sim \Lap(cb)$ for $c > 0$. Hence for $t \in
  [0,1/b]$, we have
  \begin{align*}
    \Pr[ Y \geq \acc ] &= \Pr[ e^{tY} \geq e^{t\acc} ] \leq
    \frac{E[e^{tY}]}{e^{t\acc}} \\
    & = e^{-t\acc} \prod_{i = 1}^k E[e^{tY_i}] = e^{-t\acc} \prod_i
    (1 - (q_ibt)^2)^{-1}  \\
    & = \exp( -t\acc - \sum_i \log (1 - (q_ibt)^2 )) \\
    & = \exp( -t\acc + \sum_i ((q_ibt)^2 + (q_ibt)^4/2 + (q_ibt)^6/3 +
    \cdots))
  \end{align*}
  where we used the Taylor series expansion (and hence need that
  $|q_ibt| < 1$). The last expression only worsens as the $q_i$'s
  increase, so the worst case is when all $q_i = 1$, when we get a bound
  of
  \begin{align}
    & \exp( -t\acc + k ((bt)^2 + (bt)^4/2 + (bt)^6/3 + \cdots)) \notag \\
    & \leq \exp( -t\acc + k ((bt)^2 + (bt)^4 + (bt)^6 + \cdots)) \notag \\
    & \leq \exp( -t\acc + k \frac{(bt)^2}{1 - (bt)^2}). \label{eq:4}
  \end{align}
  Let us set $t := \frac{\acc}{2kb^2}$. Recall that we needed the condition that
  $t \in [0,1/b]$, so let us assume that $\acc \leq kb$. This implies
  that $tb = \acc/2kb \leq 1/2$. Hence, plugging in this setting for
  $t$, and noting that $(1 - (tb)^2) \geq 3/4$, we get
  \begin{gather*}
    \Pr[ Y \geq \acc ] \leq \exp\left( - \frac{\acc^2}{2kb^2} + k
      \frac{\acc^2}{(3/4) (2kb)^2} \right) = \exp\left( -
      \frac{\acc^2}{6kb^2} \right).
  \end{gather*}
  This completes the proof for the case $\acc \leq kb$.  Now suppose
  $\acc > kb$; in that case let us set $t = 1/2b$---substituting this
  into~(\ref{eq:4}) gives us a tail bound of $\exp(- \acc/2b +
  k/3)$. And since $\acc > kb$, this is bounded by $\exp(-
  \acc/6b)$. This proves the theorem.
\end{proof}
}


%% file: improvedoffline.tex
\section{Towards Improving on Randomized Response for Synthetic Data}
\label{sec:improvedoffline}

In this section, we consider one possible avenue towards giving an efficient algorithm for privately generating synthetic data for graph cuts that improves over randomized response. We first show how generically, any efficient \IDC\  algorithm can be used to give an efficient \emph{offline} algorithm for privately releasing synthetic data when paired with an efficient \emph{distinguisher}. The analysis here follows the analysis of \cite{GHRU11}, who analyzed the corresponding algorithm when instantiated with the multiplicative weights algorithm, rather than a generic \IDC\  algorithm.

We will pair an \IDC\ algorithm for a class of queries $\cC$ with a corresponding \emph{distinguisher}.
\begin{definition}[$(F(\epsilon),\gamma)$-Private Distinguisher]
Let $\queryset$ be a set of queries, let $\gamma \geq 0$ and let $F(\epsilon):\mathbb{R}^+\rightarrow \mathbb{Z}$ be a function. An algorithm $\textrm{Distinguish}_\epsilon:\dbs \times \dbs \rightarrow \queryset$ is an $(F(\epsilon),\gamma)$-Private Distinguisher for $\queryset$ if for every setting of the privacy parameter $\epsilon$, it is $\epsilon$-differentially private with respect to $\db$ and if for every $\db, \db' \in \dbs$ it outputs a $\query^* \in \queryset$ such that $|\query^*(\db)-\query^*(\db')| \geq \max_{\query\in \queryset}|\query(\db)-\query(\db')| - F(\epsilon)$ with probability at least $1-\gamma$.
\end{definition}
\full{Note that in \cite{GHRU11}, we referred to a distinguisher as an agnostic learner. Indeed, a distinguisher is solving the agnostic learning problem for its corresponding set of queries. We here refer to it as a distinguisher to emphasize its applicability beyond the typical realm of learning (e.g. we here hope to apply a distinguisher to a graph cuts problem).}
\short{We present the algorithm in the full version, but the idea is very simple. Rather than waiting for a query to arrive online that induces an update step, we find queries which will induce update steps using the distinguisher. The IDC algorithm will guarantee that there will not be too many update steps, and so an efficient distinguisher will yield an efficient algorithm for releasing synthetic data.}
\full{
\begin{algorithm}
\caption{The Iterative Construction (IC) Mechanism. It takes as input an $(F(\epsilon),\gamma)$-Private Distinguisher $\textrm{Distinguish}_\epsilon$ for $\queryset$, together with an $\maxupdates(\acc)$-iterative database construction algorithm $\update_\acc$. for $\queryset$}
\label{alg:IC}
\textbf{IC}($\db,\eps,\delta,\acc,\textrm{Distinguish}, \update$):
\begin{algorithmic}
\STATE \textbf{Let} $D^0 = \update(\emptyset,\cdot,\cdot)$.
\STATE \textbf{Let} $\epsilon_0 = \epsilon_0(\acc) = \leftarrow \frac{\epsilon}{4\sqrt{\maxupdates(\acc)\log(1/\delta)}}$
\FOR{$t = 1$ to $\maxupdates(\acc)$}
 \STATE \textbf{Let} $\query^{(t)} = \textrm{Distinguish}_{\epsilon_0}(\db, D^{t-1})$
 \STATE \textbf{Let} $\noisyreal^{(t)} = \query^{(t)}(\db) + \Lap\left(\frac{1}{\eps_0}\right)$.
 \IF{$|\noisyreal^{(t)} - \query^{(t)}(D^{t-1})| < 3\acc/4$}
    \STATE \textbf{Output} $\db' = D^{t-1}$.
 \ELSE
    \STATE \textbf{Let} $D^t = \update_{\acc/2}(\db^{t-1},\query^{(t)},\noisyreal^{(t)})$.
 \ENDIF
\ENDFOR
\STATE \textbf{Output} $\db' = D^{\maxupdates(\acc)}$.
\end{algorithmic}
\end{algorithm}

What follows is a formal analysis, but the intuition for the mechanism is simple: we simply run the iterative database construction algorithm to construct a hypothesis that approximately matches $\db$ with respect to the queries $\cC$. If our distinguisher succeeds in finding a query that has high discrepancy between the hypothesis database and the true database whenever one exists, then our IDC algorithm will output a database that is $\beta$-accurate with respect to $\cC$. This requires at most $T$ iterations, and so we access the data only $2T$ times using $(\epsilon_0,0)$-differentially private methods (running the given distinguisher, and then checking its answer with the Laplace mechanism). Privacy will therefore follow from the composition theorem.

\begin{theorem}
\label{thm:IC-Privacy}
Given parameters $\epsilon,\delta < 1$, The IC mechanism is $(\eps,\delta)$ differentially private.
\end{theorem}
\begin{proof}
The mechanism accesses the data at most $2\maxupdates(\acc)$ times using algorithms that are $\eps_0$-differentially private. By Theorem \ref{thm:eps-delta-composition}, the mechanism is therefore $(\eps',\delta)$-differentially private for $\eps' = \sqrt{4\maxupdates(\acc)\ln(1/\delta)}\eps_0 + 2\maxupdates(\acc)\eps_0(e^{\eps_0}-1)$. Plugging in our choice of $\epsilon_0$ proves the claim.
\end{proof}

\begin{theorem}
\label{thm:IC-Utility}
Given an $(F(\epsilon),\gamma)$-private distinguisher and a $\maxupdates(\acc)$-IDC, the Iterative Construction mechanism is $\acc,\beta$ accurate for:
$$\acc \geq \max\left[\frac{16\sqrt{\maxupdates(\acc)\log(1/\delta)}\log(2\maxupdates(\acc)/\beta)}{\epsilon}, 2F\left(\frac{\epsilon}{4\sqrt{\maxupdates(\acc)\log(1/\delta)}}\right)\right]$$
so long as $\gamma \leq \beta/(2\maxupdates(\acc))$.
\end{theorem}

\begin{proof}
The analysis is straightforward. First we observe that because the algorithm runs for at most $\maxupdates(\acc)$ steps, except with probability at most $\beta/2$, for all $t$:
$$|\noisyreal^{(t)} -  \query^{(t)}(\db)| \leq \frac{1}{\eps_0}\log\frac{2\maxupdates(\acc)}{\beta} = \frac{4\sqrt{\maxupdates(\acc)\log(1/\delta)}}{\epsilon}\log\frac{2\maxupdates(\acc)}{\beta} \leq \frac{\acc}{4}$$
Note that by assumption, $\gamma \leq \beta/(2\maxupdates(\acc))$, so we also have that except with probability $\beta/2$,
$$|\query^{(t)}(\db) - \query^{(t)}(\db^{t-1})| \geq  \max_{\query'\in \queryset}|\query'(\db) - \query'(\db^{t-1})| - F(\frac{\epsilon}{4\sqrt{\maxupdates(\acc)\log(1/\delta)}}) \geq \max_{\query'\in \queryset}|\query'(\db) - \query'(\db^{t-1})| - \frac{\acc}{2}$$
For the rest of the argument, we will condition on both of these events occurring, which is the case except with probability $\beta$. There are two cases. Either a database $\db' = D^{\maxupdates(\acc)}$ is output, or database $\db' = D^{t-1}$ for $t \leq \maxupdates(\acc)$ is output. First, suppose $\db' = D^{\maxupdates(\acc)}$. Since for all $t$ $|\noisyreal^{(t)} - \query^{(t)}(D^{t-1})| \geq 3\acc/4$ and by our conditioning, $|\noisyreal^{(t)} -  \query^{(t)}(\db)|\leq \frac{\acc}{4}$, the sequence $(D^t,\query^{(t)},\noisyreal^{(t)})$, formed a maximal $(\update_{\acc/2},\db,\queryset,\acc/2,\maxupdates(\acc))$-\DUS. Therefore, we have that $\max_{\query \in \queryset}|\query(\db)-\query(\db')| \leq \acc/2$ as desired. Next, suppose $\db' = D^{t-1}$ for $t \leq \maxupdates(\acc)$. Then it must have been the case that for some $t$, $|\noisyreal^{(t)} - \query^{(t)}(D^{t-1})| < 3\acc/4$. By our conditioning, in this case it must be that $\query^{(t)}(\db)-\query^{(t)}(D^{t-1}) < \acc/2$, and that therefore by the properties of an $(F(\epsilon_0),\gamma)$-distinguisher:
$$\max_{\query \in \queryset}|\query(\db)-\query(\db')| < \acc/2 + F(\epsilon_0) \leq \acc$$
as desired.
\end{proof}
}
\short{

\begin{theorem}
There is an $(\eps,\delta)$-differentially private mechanism for releasing synthetic data such that given an $(F(\epsilon),\gamma)$-private distinguisher and a $\maxupdates(\acc)$-IDC, it is $(\acc,\beta)$-accurate for:
$$\acc \geq \max\left[\frac{16\sqrt{\maxupdates(\acc)\log(1/\delta)}\log(2\maxupdates(\acc)/\beta)}{\epsilon}, 2F\left(\frac{\epsilon}{4\sqrt{\maxupdates(\acc)\log(1/\delta)}}\right)\right]$$ as long as $\gamma \leq \beta/(2\maxupdates(\acc))$
\end{theorem}
We defer the proof until the full version.
}
Note that the running time of the algorithm is dominated by the running time of the IDC algorithm and of the distinguishing algorithm: efficient IDC algorithms paired with efficient distinguishing algorithms for a class of queries $\queryset$ automatically correspond to efficient algorithms for privately releasing synthetic data useful for $\queryset$. For the class of graph cut queries, both the multiplicative weights IDC and the Frieze/Kannan IDC are computationally efficient. Therefore, one approach to finding a computationally efficient algorithm for releasing synthetic data useful for cut queries is to find an efficient private distinguishing algorithm for cut queries.

One curious aspect of this approach is that it might in fact be computationally easier to release a \emph{larger} class of queries than cut queries, even though this is a strictly more difficult task from an information theoretic perspective. For example, solving the distinguishing problem for cut queries on graphs $\db$ and $\db'$ is equivalent to finding a pair of sets $(S,T)$ which witness the cut-norm on the graph $\db - \db'$. On the other hand, solving the distinguishing problem for rank-1 queries (which include cut queries, and are a larger class) is equivalent to finding the best rank-1 approximation to the adjacency matrix $\db - \db'$. The former problem is NP-hard, whereas the latter problem can be quickly solved non-privately using the singular value decomposition.

\begin{corollary}
An efficient $(F(\epsilon),\gamma)$-distinguisher for the class of rank-1 queries for $F(\epsilon) = T/\epsilon$ would yield an $(\acc,\beta)$-accurate mechanism for releasing synthetic data for graph cuts (and all rank-1 queries) for any $\beta \geq \Omega(\exp(-\epsilon T))$ and:
$\acc_{MW} =  2\sqrt[4]{2}\epsilon^{-1/2}\sqrt{Tm}\left(\log |V|\log (1/\delta)\right)^{1/4}$
using the multiplicative weights IDC, or:
$\acc_{FK} \geq 2 \epsilon^{-1/2}(m\log (1/\delta))^{1/4}\sqrt{|V| T}$
using the Frieze/Kannan IDC
\end{corollary}
\short{The proof, deferred to the full version, only requires plugging in the parameters for these two IDC algorithms.}
\full{
\begin{proof}
The Multiplicative Weights mechanism is a $\maxupdates(\acc)$-IDC for the class of rank-1 queries on graphs with $m = |E|$ edges and $|V|$ vertices for $\maxupdates(\acc) = 2m^2\log |V|/\acc^2$. We can set:
$$\acc \geq F\left(\frac{\epsilon}{4\sqrt{\maxupdates(\acc)\log(1/\delta)}}\right) = \frac{4\sqrt{2}Tm\sqrt{\log |V|\log(1/\delta)}}{\acc\epsilon}$$
which allows us to take
$$\acc \geq \frac{2\sqrt[4]{2}\sqrt{Tm}\left(\log |V|\log \frac{1}{\delta}\right)^{1/4}}{\sqrt{\epsilon}}$$
The Frieze/Kannan algorithm is a $\maxupdates(\acc)$-IDC with $\maxupdates(\acc) = m|V|^2/\acc^2$. We can set:
$$\acc \geq F\left(\frac{\epsilon}{4\sqrt{\maxupdates(\acc)\log(1/\delta)}}\right) = \frac{4T\sqrt{m\log(1/\delta)}|V|}{\acc\epsilon}$$
which allows us to take:
$$\acc \geq \frac{2(m\log 1/\delta)^{1/4}\sqrt{|V|\cdot T}}{\sqrt{\epsilon}}$$
\end{proof}
}
We remark that for the class of rank-1 queries, an efficient
$(F(\epsilon),\gamma)$-distinguisher with $F(\epsilon) =
\tilde{O}\left(\frac{|V|}{\epsilon}\right)$ would be sufficient to yield
an efficient algorithm for releasing synthetic data useful for cut
queries, with guarantees matching those of the best known algorithms for
the interactive case, as listed in Table \ref{tab:results}. For graphs
for which the size of the edge set $m \leq \Omega(n^2)$, this would
yield an improvement over our randomized response mechanism, which is
the best mechanism currently for privately releasing synthetic data for
graph cuts. We note that a distinguisher for rank-1 queries must simply
give a good rank-1 approximation to the matrix $\db - \db'$. We further
note that in the case of the Frieze/Kannan IDC for graph cuts, $\db -
\db'$ is always a symmetric matrix (because both the hypothesis is at
every step simply the adjacency matrix for an undirected graph, as of
course is the private database), and hence an algorithm for finding
accurate rank-1 approximations merely for symmetric matrices would
already yield an algorithm for releasing synthetic data for cuts!
Unlike classes of queries like conjunctions, for which their are
imposing barriers to privately outputting useful synthetic data
\cite{UV11,GHRU11}, there are as far as we know no such barriers to
improving our randomized-response based results for synthetic data for
graph cuts.  We leave finding such an algorithm, for privately giving
low rank approximations to matrices, as an intriguing open problem. 


%% file: appendix.tex
\appendix
\section{Other \IDC \ Algorithms}
\label{app:other-IDCs}
In this section, we demonstrate how the median mechanism and the multiplicative weights mechanism fit into the IDC framework. These mechanisms apply to general classes of linear queries~$\queryset$.
\subsection{The Median Mechanism}
In this section, we show how to use the median database subroutine as an \IDC.
\begin{definition}[Median Datastructure]
A median datastructure $\mathbf{D}$ is a collection of databases $\mathbf{D} \subset \dbs$. Any query can be evaluated on a median datastructure as follows: $\query(\mathbf{D}) = \textrm{Median}(\{\query(\db') : \db' \in \mathbf{D}\})$.
\end{definition}

\begin{algorithm} \label{alg:MM}
\caption{The Median Mechanism (MM) Algorithm.}
\noindent\textbf{$\update^{MM}_{k,\acc}(\mathbf{D}^t, \querystep{t}, \noisyrealstep{t})$}
\begin{algorithmic}
\STATE{\textbf{If:} $\mathbf{D}^t = \emptyset$ \textbf{then:} output $\mathbf{D}^0 = \{\db \in \dbs : |\db| = n^2 \log \queries /\acc^2\}$}
\STATE{\textbf{Else if:} $\querystep{t}(\mathbf{D}^t) - \noisyrealstep{t} > 0$ \textbf{then:} output $\mathbf{D}' = \mathbf{D}' \setminus \{\db \in \mathbf{D} : \querystep{t}(\db) \geq \querystep{t}(\mathbf{D})\} $}
\STATE{\textbf{Else if:} $\querystep{t}(\mathbf{D}^t) - \noisyrealstep{t} < 0$ \textbf{then:} output $\mathbf{D}' = \mathbf{D}' \setminus \{\db \in \mathbf{D} : \querystep{t}(\db) \leq \querystep{t}(\mathbf{D})\} $}
\end{algorithmic}
\end{algorithm}

\begin{theorem}
\label{thm:MM-Decomposition}
The Median Mechanism algorithm is a $\maxupdates(\acc) = n^2 \log |\univ|\log \queries/\acc^2$ iterative database construction algorithm for every class of $\queries$ linear queries $\queryset$.
\end{theorem}
\begin{proof}
Let $\db \in \dbs$ be any database and let $\set{(\mathbf{D}^t, \querystep{t}, \noisyrealstep{t})}_{t=1,\dots, \maxupdates}$ be a $(\update^{MM}_k, \mathbf{D}^*, \queryset, \acc, \maxupdates)$-\dus.  We want to show that $\maxupdates(\acc) \leq  n^2 \log |\univ|\log k/\acc^2$.  Specifically, that after $ n^2 \log |\univ|\log \queries/\acc^2$ invocations of $\update^{MM}_{k,\acc}$, the median datastructure $\mathbf{D}^{n^2 \log |\univ|\log \queries/\acc^2}$ is $(\acc, \queryset)$-accurate for $\db$. The argument is simple. First, we have a simple fact from \cite{BLR08}:
\begin{claim}
For any set of $\queries$ linear queries $\queryset$ and any database $\db$ of size $n$, there is a database $\db'$ of size $|\db'| = n^2 \log k/\acc^2$ so that $\db'$ is $\acc$-accurate for $\db$ with respect to $\queryset$.
\end{claim}
From this claim, we have that $|\mathbf{D}^t| \geq 1$ for all $t$, and so can always be used to evaluate queries. On the other hand, each update step eliminates half of the databases in the median datastructure: $|\mathbf{D}^t| = |\mathbf{D}^{t-1}|/2$. This is because the update step eliminates every database either above or below the median with respect to the last query. Initially $|\mathbf{D}^0| = |\univ|^{n^2 \log \queries/\acc^2}$, and so there can be at most $\maxupdates(\acc) \leq \log n^2 |\univ|\log \queries/\acc^2$ update steps before we would have $|\mathbf{D}^\maxupdates| < 1$, a contradiction.
\end{proof}

\subsection{The Multiplicative Weights Mechanism}
In this section we show how to use the multiplicative weights subroutine as an \IDC. The analysis of the multiplicative weights algorithm is not new, and follows \cite{HR10}. It will be convenient to think of our databases in this section as probability distributions, i.e. normalized so that $||\db||_1 = 1$. Note that if we are $\acc/n$ accurate for the normalized database, we are $\acc$-accurate for the un-normalized database with respect to any set of linear queries.

\begin{algorithm}
\caption{The Multiplicative Weights (MW) Algorithm.}
\label{alg:MW}
$\update_\acc^{MW}(\db^t,\querystep{t}, \noisyrealstep{t}$):
\begin{algorithmic}
\STATE \textbf{Let} $\eta \leftarrow \acc/(2n)$. 
\STATE{\textbf{If:} $\db^t = \emptyset$ \textbf{then:} output $\db' = \db \in  \reals^{|\univ|}$ such that $D^0_i = 1/|\univ|$ for all $i$.}
\IF{$\noisyrealstep{t}  <  \querystep{t}(\db^t)$}
    \STATE \textbf{Let} $r_t = \querystep{t}$
\ELSE
    \STATE \textbf{Let} $r_t = 1-\querystep{t}$
\ENDIF
\STATE \textbf{Update:} For all $i \in [|\univ|]$ Let $$\hat{\db}^{t+1}_i = \exp(-\eta r_t(\db^t_i))\cdot \db^t_i$$ $$\db^{t+1}_i = \frac{\hat{\db}^{t+1}_i}{\sum_{j=1}^{|\univ|}\hat{\db}^{t+1}_j}$$
\STATE \textbf{Output} $\db^{t+1}$.
\end{algorithmic}
\end{algorithm}

\begin{theorem}
\label{thm:MW-Decomposition}
The Multiplicative Weights algorithm is a $\maxupdates(\acc) =  4 n^2 \log |\univ|/\acc^2$ iterative database construction algorithm for every class of linear queries $\queryset$.
\end{theorem}
\begin{proof}
Let $\db \in \dbs$ be any database and let $\set{(\dbstep{t}, \querystep{t}, \noisyrealstep{t})}_{t=1,\dots, \maxupdates}$ be a $(\update^{MW}, \db^*, \queryset, \acc, \maxupdates)$-\dus.  We want to show that $\maxupdates(\acc) \leq  4n^2 \log |\univ|/\acc^2$.  Specifically, that after $ 4n^2 \log |\univ|/\acc^2$ invocations of $\update^{MW}$, the database $\dbstep{4n^2 \log |\univ|/\acc^2}$ is $(\acc, \queryset)$-accurate for $\db$.
First let $\hat{\db} \in \reals^{|X|}$ be a normalization of the database $\db$: $\hat{\db}_i = \db_i/\|\db\|_1$. Note that for any linear query, $\query(\db) = n\cdot \query(\hat{\db})$. We define:
$$\Psi_t \eqdef D(\hat{\db}||D^t) = \sum_{i=1}^{|\univ|}\hat{\db}_i\log\left(\frac{\hat{\db}_i}{D^t_i}\right)$$
We begin with a simple fact:
\begin{claim}[\cite{HR10}]
For all $t$: $\Psi_t \geq 0$, and $\Psi_0 \leq \log|\univ|$.
\end{claim}
We will argue that in every step for which $|\querystep{t}(\db) - \querystep{t}(\db^t)| \geq \acc/n$ the potential drops by at least $\acc^2/4n$. Because the potential begins at $\log |\univ|$, and must always be non-negative, we know that there can be at most $\maxupdates(\acc) \leq 4n^2\log|X|/\acc^2$ steps before the algorithm outputs a database $\db^t$ such that $\max_{\query\in \queryset}|\query(\db) - \query(\db^t)| < \acc/n$, which is exactly the condition that we want.
\begin{lemma}[\cite{HR10}]
\label{lem:MWPotentialDrop}
$$\Psi_{t}-\Psi_{t+1} \geq \eta\left(r_t(\db^{t}) -  r_t(\db)\right) - \eta^2$$
\end{lemma}
\begin{proof}
\begin{eqnarray*}
\Psi_{t}-\Psi_{t+1} &=& \sum_{i=1}^{|\univ|}\hat{\db}_i\log\left(\frac{D^{t+1}_i}{D^{t}_i}\right) \\
&=& -\eta r_t(\db) -\log\left(\sum_{i=1}^{|\univ|}\exp(-\eta r_t(x_i))D^{t}_i\right) \\
&\geq& -\eta r_t(\db) - \log\left(\sum_{i=1}^{|\univ|}D^{t}_i(1+\eta^2-\eta r_t(x_i))\right) \\
&\geq& \eta\left(r_t(\db^{t})  -  r_t(\db)\right) - \eta^2
\end{eqnarray*}
\end{proof}
The rest of the proof now follows easily. By the conditions of an iterative database construction algorithm, $|\noisyrealstep{t} - \querystep{t}(\db)| \leq \acc/(2n)$. Hence, for each $t$ such that $|\querystep{t}(\db) - \querystep{t}(\db^t)| \geq \acc/n$, we also have that $\querystep{t}(\db) > \querystep{t}(\db_t)$ if and only if $\noisyrealstep{t} > \querystep{t}(\db_t)$. In particular, $r_t = \querystep{t}$ if $\querystep{t}(\db^t) - \querystep{t}(\db) \geq \acc/n$, and $r_t = 1-\querystep{t}$ if $\querystep{t}(\db)-\querystep{t}(\db^t) \geq \acc/n$. Therefore, by Lemma \ref{lem:MWPotentialDrop} and the fact that $\eta = \acc/2n$:
$$\Psi_{t}-\Psi_{t+1} \geq \frac{\acc}{2n}\left(r_t(\db^{t}) -  r_t(\db)\right) - \frac{\acc^2}{4n^2} \geq \frac{\acc}{2n}\left(\frac{\acc}{n} \right) - \frac{\acc^2}{4n^2} = \frac{\acc^2}{4n^2}$$
\end{proof}
